\documentclass{article}
\usepackage{emulateapj}
\usepackage{onecolfloat}
\usepackage{graphicx} 

\lefthead{Bernstein, Freedman, \& Madore}
\righthead{Detections of the Optical EBL (III): Cosmological Implications}

\def\BFM{Bernstein, Freedman, \& Madore}
\def\ABmag{AB\,mag}
\def\hi{\noindent \hangindent=2.5em}
\def\etal{{\it et\thinspace al.\ }}

\def\slantfrac#1#2{\hbox{$\,^#1\!/_#2$}}

\def\ABmag{AB\,mag}

\def\ion#1#2{\rm{#1}{\sc{#2}}\relax}
\def\nodata{ ~$\cdots$~ }
\def\tto#1{$\times10^{#1}$}

\def\escsa{ergs s$^{-1}$ cm$^{-2}$ sr$^{-1}$ \AA$^{-1}$}
\def\escsh{ergs s$^{-1}$ cm$^{-2}$ sr$^{-1}$ Hz$^{-1}$}

\def\esa{ergs s$^{-1}$ \AA$^{-1}$}
\def\tto#1{$\times10^{#1}$}

\def\micron{\hbox{$\mu$m}}

\def\farcm{\hbox{.\kern -0.7ex\raisebox{.9ex}{\scriptsize$\prime$}}}
\def\farcs{\hbox{\kern 0.13ex.\kern -0.95ex%
  \raisebox{.9ex}{\scriptsize$\prime\prime$}\kern -0.1ex}}

\def\spose#1{\hbox to 0pt{#1\hss}}
\def\lesssim{\mathrel{\hbox{\rlap{\hbox{\lower4pt\hbox{$\sim$}}}\hbox{$<$}}}}
\def\gtrsim{\mathrel{\hbox{\rlap{\hbox{\lower4pt\hbox{$\sim$}}}\hbox{$>$}}}}
\def\lta{\mathrel{\spose{\lower 3pt\hbox{$\mathchar"218$}}
     \raise 2.0pt\hbox{$\mathchar"13C$}}}
\def\gta{\mathrel{\spose{\lower 3pt\hbox{$\mathchar"218$}}
     \raise 2.0pt\hbox{$\mathchar"13E$}}}

\submitted{Submitted: 13 June 2000 ; 
Revised:  11 August 2001}

\begin{document}

\twocolumn[
 
\title{ The First Detections of the Extragalactic Background
	Light at 3000, 5500, and 8000\AA\ (III):
        Cosmological Implications}
 
\author{Rebecca A. Bernstein\altaffilmark{1,2,3}}
\author{Wendy L. Freedman\altaffilmark{2}}
\author{Barry F. Madore\altaffilmark{2,4}}
 
\affil{\footnotesize 1) Division of Math, Physics, and Astronomy,
		California Institute of Technology,
                Pasadena, CA 91125}
\affil{\footnotesize 2) Carnegie Observatories,
                813 Santa Barbara St, 
                Pasadena, CA 91101}
\affil{\footnotesize 3) rab@ociw.edu, Hubble Fellow}
\affil{\footnotesize 4) NASA/IPAC Extragalactic Database, 
                California Institute of Technology, 
                Pasadena, CA 91125}

\begin{abstract}
We have used the Hubble Space Telescope Wide--Field and Planetary
Camera 2, in combination with ground--based spectroscopy, to measure
the integrated flux of galaxies at optical wavelengths --- the
extragalactic background light (EBL).  We have also computed the
integrated light from individual galaxy counts in the images used to
measure the EBL and in the Hubble Deep Field. We find that flux in
galaxies as measured by standard galaxy photometry methods has
generally been underestimated by about 50\%, resulting from missed
flux in the outer, lower surface--brightness parts of galaxies and
from associated errors in the estimated sky level.  Comparing the
corrected, integrated flux from individual galaxies with our total EBL
measurement, we find that there is yet further light that contributes
to the background that is not represented by galaxy counts, and that
the total flux in individually detected sources is a factor of 2 to 3
less than the EBL from 8000 to 3000 \AA.  We show that a significant
fraction of the EBL may come from normal galaxies at $z<4$, which are
simply undetectable as a result of K-corrections and cosmological
surface brightness dimming.  This result is consistent with results
from recent redshift surveys at $z<4$.  In the context of some simple
models, we discuss the constraints placed by the EBL on evolution in
the luminosity density at $z>1$; while significant flux comes from
galaxies beyond the current detection limits, this evolution cannot
be tightly constrained by our data.

Based on our measurements of the optical EBL, combined with previously
published measurements of the UV and IR EBL, we estimate that the
total EBL from 0.1--1000 \micron\ is 100$\pm$20 nW m$^{-2}$ sr$^{-1}$.
If the total EBL were produced entirely by stellar nucleosynthesis,
then we estimate that the total baryonic mass processed through stars
is $\Omega_\ast = 0.0062 (\pm 0.0022) h^{-2}$ in units of the critical
density. For currently favored values of the baryon density,
$\Omega_{\rm B}$, this corresponds to 
$0.33\pm0.12 \Omega_{\rm B}$. 
This estimate is smaller by roughly 7\% if we allow for a
contribution of 7$h_{0.7}$ nW m$^{-2}$ sr$^{-1}$ to the total EBL from
accretion onto central black holes.  This estimate of $\Omega_*$
suggests that the universe has been enriched to a total metal mass of
$0.21(\pm0.13) Z_\odot \Omega_{\rm B}$, which is consistent with
other observational estimates of the cumulative metal mass fraction of
stars, stellar remnants, and the intracluster medium of galaxy
clusters in the local universe.

\end{abstract}

%------------------- ------------------- ------------------- --------------


%------------------- ------------------- ------------------- --------------

]

\setcounter{footnote}{0}

\section{Introduction} \label{intro}
 
The integrated optical flux from all extragalactic sources is a record
of the stellar nucleosynthesis in the universe and the chemical
evolution which has resulted from it.  In \BFM\ 2001a (henceforth,
Paper I), we presented detections of the optical EBL in the HST/WFPC2
wide-band filters F300W ($U_{300}$, $\lambda_0\sim$ 3000\AA), F555W
($V_{555}$, $\lambda_0\sim$ 5500\AA), and F814W ($I_{814}$,
$\lambda_0\sim$ 8000\AA) based on simultaneous data sets from Hubble
Space Telescope (HST) and Las Campanas Observatory (LCO).  In \BFM\
2001b (henceforth, Paper II), we presented details of a measurement of
the diffuse foreground zodiacal light which we use in Paper I.  Here
we briefly summarize the results of Papers I and II and discuss the
cosmological implications of these detections of the EBL.

The majority of the EBL at UV to IR wavelengths is produced by stars
at restframe wavelengths of 0.1--10\micron.  Due to cosmic expansion,
the EBL at $U_{300}$, $V_{555}$, and $I_{814}$ potentially includes
redshifted light from stellar populations out to $z\sim8$ (the
redshifted Lyman--limit cut--off of the $I_{814}$ filter).  Although
stars themselves do not emit much light at wavelengths longer than
10\micron, a complete census of the energy produced by stellar
nucleosynthesis in the universe must consider the EBL over the full
wavelength range 0.1-1000\micron\ because dust in the emitting
galaxies will absorb and re-radiate starlight, redistributing energy
from nucleosynthesis into the thermal IR.

With 8m--class telescopes and HST, the limits of resolved--source
methods (i.e., number counts, redshift surveys, QSO absorption lines,
etc.) for studying star formation in the universe are being extended
to ever fainter levels; however, a direct measurement of the EBL
remains an invaluable complement to these methods.  Galaxies with low
{\it apparent} surface brightness --- both intrinsically low surface
brightness galaxies at low redshift and normal surface brightness
galaxies at high redshift --- are easily missed in
surface--brightness--limited galaxy counts and consequentally in
follow--up redshift surveys.  Identification, not to mention
photometry, of faint galaxies becomes very uncertain near the
detection limits.  Even efforts to understand galaxy evolution,
chemical enrichment, and star formation through QSO absorption line
studies appear to be biased against chemically enriched, dustier
systems, as these systems can obscure QSOs which might lie behind
them (Fall \& Pei 1989, Pei \& Fall 1995, Pettini \etal 1999).  In
contrast, a direct measurement of the spectral energy distribution
(SED) of the EBL from the UV to the far--IR is a complete record of
the energy produced by star formation and is immune to surface
brightness selection effects.

In addition to energy originating from stellar nucleosynthesis, the
EBL includes energy emitted by accreting black holes in quasars and
active galactic nuclei.  However, at optical wavelengths, the quasar
luminosity functions at redshifts $z\lesssim5$ indicate that the
optical luminosity density from quasars is a small fraction
($\sim2.5$\%) of the that from galaxies (e.g. Boyle \& Terlevich
1998).  In addition, our measurement of the EBL excludes any point--like
sources (of which there are 3 in our images), under the prior
assumption that those sources are Galactic foreground stars.  We
therefore expect quasars to be a negligible source of flux in our
measurements of the optical EBL.

The contribution from active galactic nuclei (AGN) is more difficult
to assess, as recent dynamical evidence (Richstone \etal  1998)
indicates that massive black holes reside in most galaxies and
sensitive optical spectroscopy (Ho \etal  1997a, 1997b) indicates
that AGN have at least a weak contribution to more than 50\% of nearby
galaxies. Nonetheless, simple accretion models, the total X-ray
background, and the X-ray to far--IR spectral energy distribution of
AGN and quasars all indicate that the total contribution to the
bolometric EBL from accretion onto central black holes is $\lta15$\%
(see \S\ref{agn}), and is emitted at thermal IR wavelengths.  In
principle, measurements of the EBL also constrain possible the total
energy output from more exotic sources, such as gravitationally
collapsing systems, brown dwarfs, and decaying particles (see Carr,
Bond, \& Hogan 1986, 1991 and Dwek \etal  1998 for discussions).

The outline of the paper is as follows. In \S\ref{sum.obs}, we give an
overview of the observations and methods used to measure the EBL as
discussed in Papers I and II.  In \S\ref{sum.detect}, we summarize the
individual measurements and associated errors we have obtained from
each data set and the final EBL detections which result from them.  In
\S\ref{opt.res}, we compare the measured EBL with the integrated
optical flux from resolvable sources as quantified by number counts
and luminosity functions. In \S\ref{opt.unres}, we quantify the
contributions to the optical EBL which one might expected from sources
which fall below the detection limits of the HDF based on explicit
assumptions regarding the surface brightness, luminosity, and redshift
distribution of galaxy populations in the universe.  In \S\ref{bolom},
we discuss models of the SED of the EBL based on these and recent
results in the far infrared. Finally, in \S\ref{stel} we discuss the
total star formation and chemical enrichment history of the universe
required to produce the bolometric flux of the EBL, and compare
the inferred values to other observations of the total baryon fraction
in stars and the metal mass density in the local universe.
We abbreviate the adopted units \escsa\  as cgs throughout.

\section{Summary of Observations}\label{sum.obs}

As is true of all background measurements, the difficulty in measuring
the optical EBL is in differentiating it from the much brighter
foregrounds: terrestrial airglow, zodiacal light (ZL), and diffuse
Galactic light (DGL).  Relative to the EBL flux at $\sim$5000\AA,
airglow and ZL are each more than 100 times brighter than the
EBL. Along the most favorable lines of sight, the DGL is roughly equal
in flux to the EBL.  We have measured the EBL in a field which is out
of the ecliptic plane and near the Galactic pole in order to optimally
minimize the contributions of zodiacal light, DGL, and nearby stars
(see Paper I).

In the EBL measurement presented in Paper I, we have used three
simultaneous data sets to isolate the diffuse EBL from the foreground
sources: (1) absolute surface photometry taken with WFPC2 aboard HST
using the wide--band filters F300W ($U_{300}$), F555W ($V_{555}$), and
F814W ($I_{814}$); (2) low resolution ($\sim300$\AA) surface
spectrophotometry at 4000---7000\AA\ taken with the FOS, also aboard
HST; and (3) moderate resolution ($\sim$2\AA) surface
spectrophotometry taken with the Boller and Chivens spectrograph on
the 2.5m duPont telescope at LCO.  We use the two HST data sets to
measure the total mean flux of the night sky, including ZL, DGL and
the EBL. We avoid terrestrial airglow all together by using HST for
this measurement.  We then use the LCO spectra to measure the absolute
surface brightness of ZL in the same field and on the same nights as
the HST observations.  Finally, we estimate the small DGL contribution
using a scattering model which is in good agreement with the
observations.  We then subtract the measured ZL and the modeled DGL
from the total flux measured with HST/WFPC2 through each filter and
with HST/FOS.  These measurements are
described in detail in Papers I and II.  Below, we summarize the
observations, results, and accuracy of the individual measurements
which contribute to the EBL detection (see Table
\ref{tab:total.summary}). 

Bright galaxies brighter are not statistically
well sampled in the 4.4 arcmin$^2$ WFPC2 field of view.  We have,
therefore, masked out any sources brighter than $V_{555}=23$\ABmag\ in
the WFPC2 images before we measured the total sky flux. To do so, we
used masks which are derived from the F555W images and extend to
four times the isophotal radius in those data.  {\it We
use the abbreviation} EBL23 {\it as a reminder of this bright
magnitude cut--off.}  The EBL23 detections can be combined with
ground--based counts at $V_{555}\!<\!23$\ABmag\ to obtain the total
EBL.  The WFPC2 surface brightness measurements have random errors of
$<1$\% and systematic uncertainties of 1--2\% of the total background
flux.  From the HST/WFPC2 data alone, we can also identify a minimum
flux from {\it detectable} sources.  This minimum is given in Table
\ref{tab:total.summary}, and the method used to obtain this result is
summarized in \S\ref{sum.detect}.
 
The FOS spectra also provide a measurement of total flux. The random
error per resolution element is around 2.1\%, and the systematic
uncertainty over the full range is 3.5\%.  The $\sim 14$
arcsec$^2$ FOS field of view and $\sim 4$\% systematic uncertainty
make the FOS less useful than the WFPC2 for measuring the EBL.
However, most of systematic uncertainty is due to the poorly
constrained solid angle of the aperture and aperture correction. Both
of these are wavelength--independent errors, so that the FOS spectra
do provide a $\pm1$\% measurement of the color of the total
background, which is dominated by zodiacal light.

The scattering which produces the ZL is well described by classical
Mie theory for the large ($>10$\micron), rough dust grains which
populate the zodiacal dust cloud. The scattering efficiency of the
dust is only weakly wavelength dependent, so that the solar spectral
features are well preserved in the scattered ZL spectrum. However the
broad band spectrum of the zodiacal light is redder than the solar spectrum
by about 5\% per 1000\AA\ (see Paper II for further discussion) due to
surface roughness of the grains, which decreases scattering efficiency
at shorter wavelengths.  The mean ZL flux in a narrow band can thus be
measured from the apparent equivalent width of the solar Fraunhofer
lines evident in its spectrum. Small color corrections can then be
used to infer the full spectrum relative to that measurement.  This
requires moderate resolution spectra ($\sim 2$\AA) with excellent flux
calibration ($\pm 1$\%), which can only be obtained with ground--based
observations, and  then only at wavelengths where atmospheric emission
lines are relatively weak.  We have, therefore, measured the ZL in the
range 4000--5100\AA\ using spectra taken at LCO.  The resulting
measurement has a statistical error of $<1$\% and a systematic
uncertainty of $\sim 1.2$\%.  This measurement has been extrapolated the
3000\AA\ and 8000\AA\ WFPC2 bandpasses using measurements of the color
of the ZL from the FOS and ground--based LCO data.

Within the Galaxy, there is both resolved flux from discrete stars and
diffuse light (diffuse Galactic light, DGL) from starlight scattered
by interstellar dust.  Discrete stars can simply be resolved and
subtracted in the WFPC2 images.  The intensity of the dust--scattered
optical DGL and the 100\micron\ thermal emission from the same dust
are both proportional to the dust column density and the strength of
the interstellar radiation field.  To minimize the optical DGL, our
field was selected to have very low 100\micron\ emission.  The
remaining low--level DGL which does contribute has been estimated
using a simple scattering model based on the dust column density and
interstellar radiation field along the line of sight and empirical
scattering characteristics for interstellar dust.  The predictions of
this model are in good agreement with observations of the DGL from
2500--9000\AA\ (see Witt~et~al.\ 1997 and references therein).
Finally, although isotropic line emission from warm interstellar gas
is measured at all Galactic latitudes, the strongest line, H$\alpha$,
does not lie within any our HST/WFPC2 bandpasses. The next strongest
lines, [\ion{O}{iii}], are twenty times weaker and contribute
negligibly to our results.

The EBL cannot be measured in typical HST data. Our HST observations
were scheduled to avoid contaminating scattered light from all
anticipated sources: the bright Earth limb, the Moon, and off-axis stars.
Also, observations from LCO and HST were strictly simultaneous to
guarantee that the ZL measured from the ground is exactly the
contribution seen by HST.  As an additional safeguard, observations
were scheduled in 3 visits, allowing us to look for possible off--axis
scattered light with the satellite oriented at different roll angles,
to safeguard against unidentified photometric anomalies with the
instruments, and to confirm the expected modulation in the ZL with the
Earth's orbital position.

\section{Summary of EBL Detections}\label{sum.detect}

\begin{deluxetable}{l l l r r r }
\tablewidth{30pc}
\tablecaption{Summary of Measurements
\label{tab:total.summary}}
\tablehead{
\colhead{Source} &\colhead{Bandpass}&\colhead{Data source}
& \colhead{Flux} & \colhead{Random} &\colhead{Systematic }}
\startdata
Total  
 & F300W 	& {\sc hst/wfpc2}
		& 33.5		&$(\pm$ 4.9\%)	&$[\pm$ 5.6\%]\nl
\ \ Background
 & F555W 	& {\sc hst/wfpc2}
		& 105.7 	&$(\pm$ 0.3\%) 	&$[\pm$ 1.4\%]\nl
 & F814W 	&  {\sc hst/wfpc2}
		& 72.4 		&$(\pm$ 0.2\%) 	&$[\pm$ 1.4\%]\nl
 & F555W\tablenotemark{a}
	 	&  {\sc hst/fos}
		& 111.5		&$(\pm$ 0.7\%) 	&$[\pm$ 2.8\%]\nl
Zodiacal
 & 4600--4700\AA& {\sc lco} 
		& 109.4		&$(\pm$ 0.6\%)	&[$\pm$ 1.1\%]	\nl
\ \ Light
 & F300W 	& {\sc lco}\tablenotemark{b} 
		& 28.5		&$(\pm$ 0.6\%)	&[-1.1\%,+1.2\%] \nl
						%&[$\pm$ 0.6\%]	\nl
						%&[-0.7\%,+2.1\%] \nl
 & F555W 	& {\sc lco}\tablenotemark{b} 
		& 102.2 	&$(\pm$ 0.6\%)	&[-1.1\%,+1.1\%] \nl
						%&[$\pm$ 0.6\%]	\nl
						%&[-1.0\%,+1.0\%] \nl
 & F814W 	& {\sc lco}\tablenotemark{b} 
		& 69.4		&$(\pm$ 0.6\%)	&[-1.3\%,+1.1\%] \nl
						%&[$\pm$ 0.6\%]	\nl
						%&[-3.4\%,+2.0\%] \nl
%
Diffuse	
 & F300W & {\sc dgl} model & 1.0	& \nodata  	& [+25\%,-50\%] \nl
\ \ Galactic
 & F555W & {\sc dgl} model & 0.8	& \nodata 	& [+25\%,-50\%] \nl
\ \ Light
 & F814W & {\sc dgl} model & 0.8	& \nodata 	& [+25\%,-50\%] \nl
\enddata
\tablecomments{All fluxes are in units of 1\tto{-9}\escsa.  For a
source with constant flux in $ F_{\lambda}$, filters F300W, F555W, and
F814W have effective wavelengths $\lambda_0(\Delta\lambda)$
=3000(700), 5500(1200), and 8000(1500)\AA.  Individual sources of
error contributing to each measurement are summarized in Tables 3 \& 4
of Paper I and Table 1 of Paper II.}
\tablenotetext{a}{Observed FOS spectrum, convolved with the
WFPC2/F555W bandpass to allow direct comparison with the WFPC2
results.}
\tablenotetext{b}{LCO measurement of zodiacal light have been
extrapolated to the WFPC2 bandpass by applying a correction for
changing zodiacal light color with wavelength relative to the solar
spectrum.  The zodiacal light flux through the WFPC2 bandpasses was
identified using 
SYNPHOT models, the uncertainty due to which is included in the
uncertainty for the filter calibration and is shared with the
systematic uncertainty for the total background flux.}
\end{deluxetable}

\begin{deluxetable}{lcccr}
\tablewidth{30pc}
\tablecaption{EBL Results and Uncertainties
\label{tab:cum.errors}}
\tablehead{
%\tablevspace{-10pt}
\colhead{Bandpass}& 
	\colhead{Random} &
	\colhead{Systematic}& 
	\colhead{Combined}&
	\colhead{EBL($\pm1\sigma$)}\\ 
\colhead{} &
	\colhead{$\sigma_R$ (68\%)} &
	\colhead{$\sigma_S$ (68\%)} &
	\colhead{$\sigma$ (68\%)} &
	\colhead{} }
\startdata
%\tablevspace{-10pt}
\multicolumn{5}{l}{\underline{Detected EBL23 (WFPC2 + LCO)}\tablenotemark{a}} \nl
F300W\hspace{0.5in}     & 2.1	& 1.5           & 2.5	& 4.0  $(\pm 2.5)$  \nl
        F555W           & 0.6	& 1.3           & 1.4	& 2.7  $(\pm 1.4)$  \nl
        F814W           & 0.4	& 0.9		& 0.0	& 2.2  $(\pm 1.0)$  \nl
\multicolumn{5}{l}{\underline{Minimum EBL (WFPC2)}\tablenotemark{a}} \nl 
        F300W           & 0.19	& 0.13		& 0.22	& 3.2  $(\pm 0.22)$ \nl
        F555W           & 0.003	& 0.009		& 0.01	& 0.89 $(\pm 0.01)$ \nl
        F814W           & 0.002	& 0.007		& 0.01	& 0.76 $(\pm 0.01)$ \nl
\multicolumn{5}{l}{\underline{Detected EBL23 (FOS + LCO)}\tablenotemark{a}} \nl
        F555W           & 0.7	& 2.7           & 2.8	& 8.5  $(\pm 5.6)$  \nl
\multicolumn{5}{l}{\underline{Flux from detected sources 
				in HDF ($m>23$ AB mag)}} \nl
        F300W           &	&		&	& 0.66	 	\nl
        F450W           &	&		&	& 0.51	 	\nl
        F606W           &	&		&	& 0.40	 	\nl
        F814W           &	&		&	& 0.27	 	\nl
\multicolumn{5}{l}{\underline{Published number counts}\tablenotemark{b}}\nl
        \multicolumn{3}{l}{F300W ($18<U_{300}<23$ AB mag)}
						&	& 0.27 $(\pm 0.05)$ \nl
        \multicolumn{3}{l}{F555W ($15<V_{555}<23$ AB mag)}
						&	& 0.49 $(\pm 0.10)$ \nl
        \multicolumn{3}{l}{F814W ($13<I_{814}<23$ AB mag)}
						&	& 0.65 $(\pm 0.13)$ \nl
 \enddata
\tablecomments{All fluxes and errors are given in units of $10^{-9}$
\escsa.}  
\tablenotetext{a}{The systematic and statistical errors have been
combined assuming a flat or Gaussian probability distribution,
respectively,  as discussed in Paper I.  We equate
$1\sigma$  combined errors with the 68\% confidence interval, as the
combined errors are nearly Gaussian distributed. Individual sources of
error contributing to these totals are summarized in Tables 3 and 4 of
Paper I  and in  Table 1 of Paper II.}
\tablenotetext{b}{Estimated errors correspond to 
uncertainties in the fits to published galaxy counts.  The values
given correspond to $0.081\times10^{-20}$, $0.46\times10^{-20}$, and
$1.5\times10^{-20}$ in units of \escsh\ and are consistent with those
used in Pozzetti et al.\ (1998).}
\end{deluxetable}

\begin{figure}[t]
\begin{center}
\includegraphics[width=3in,angle=0]{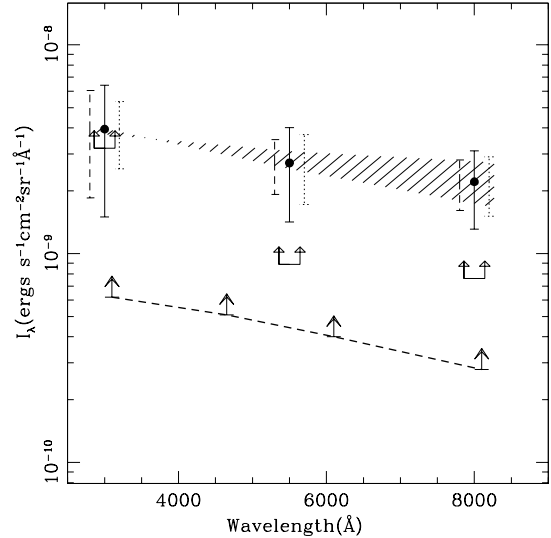}
\caption{\footnotesize Summary of EBL23 measurements, repeated from
Paper I.  Filled circles show the EBL23 obtained from surface
photometry of the total background measured from HST/WFPC2, the
zodiacal light as measured from Las Campanas Observatory, and models
of the diffuse galactic light as summarized in \S\ref{sum.detect}.
The solid, dotted, and long--dashed error bars show the combined,
systematic, and random $1\sigma$ errors, respectively.  The
hatch--marked region shows the $1\sigma$ uncertainty in the detected
EBL due to uncertainty in ZL color.  The lower limit arrows connected
by a dashed line indicate the total flux from individually
photometered galaxies with magnitudes $23<V_{555} <30$\,AB\,mag in the
HDF catalog. The u-shaped lower limit arrows show minEBL23, which is
the flux as determined by ensemble photometry from galaxies with $23
<V_{555}\leq28$\,AB\,mag in the EBL fields.}
\label{fig:eblsummary}
\end{center}
\end{figure}

The individual measurements which are combined to obtain our
detections of the EBL are summarized in Table \ref{tab:total.summary}.
We summarize our confidence intervals on the detected EBL23 in Table
\ref{tab:cum.errors} and Figure \ref{fig:eblsummary}.  For comparison
with the EBL23 fluxes, we have included in Table \ref{tab:cum.errors}
the integrated flux from individually photometered sources in the HDF,
as measured using the photometry package FOCAS (Jarvis \& Tyson 1981,
Valdes 1982) and published in Williams \etal (1996).  As the values in
this table show, the mean EBL23 detections in each bandpass are more
than $5\times$ higher than the integrated flux in HDF galaxies as
measured by standard photometry.

To help understand this large difference between the detected EBL and
the flux in HDF number counts, we have measured the total flux from
resolved galaxies in our WFPC2 images ($23<V_{555}<27.5$) using a
method which we call ``ensemble aperture photometry.''  This method is
uniquely suited to both our goal of a very accurate measurement of the
{\it ensemble} flux of all galaxies in our images and to our data set,
which has zero--point calibration accurate to $\pm 0.1\%$ over each
image and negligible scattered light. This method is described in
detail in Paper I and summarized below.

Briefly, we identify the {\it total} flux from detectable galaxies
fainter than $V_{\rm cut}=23$ \ABmag\ by simply masking out galaxies
with $V<V_{\rm cut}$ \ABmag\ (and all stars) and averaging the flux of
every pixel in the frame. From this, we obtain the mean surface
brightness of foregrounds plus {\it all} extragalactic sources, or the
average surface brightness per pixel from ``objects + sky.''  We then
mask out {\it all detected} objects using standard detection apertures
(twice the isophotal area) and calculate the average flux in the
remaining pixels. From this, we obtain the mean surface brightness
outside of the galaxy masks, or the average surface brightness per
pixel from ``sky.''  The difference between these two measurements is
then the ensemble surface brightness of all sources within the area of
the masks.  This assumes that the sky level is uniform, which is
the case to better
than 1\% accuracy in our images.  By varying the
bright magnitude cut--off ($V_{\rm cut}$) we choose for measuring
``objects + sky,'' we can isolate the flux coming from sources fainter
than $V_{\rm cut}$.

As discussed in Paper I, we find that the recovered flux increases
steadily with increasing mask size. For example, roughly 20\% of the
light from galaxies 4 magnitudes above the detection limit lies at
radii $\sqrt{2} r_{\rm iso}< r < 4 r_{\rm iso}$ (see Figure
\ref{fig:corfac}), beyond the standard--size galaxy apertures
($\sqrt{2} r_{\rm iso}$) used in faint galaxy photometry packages,
such as SExtractor (Bertin \& Arnouts 1996) or FOCAS.  Because
estimates of the sky level in standard photometry packages come from
just beyond the detection apertures, these sky estimates will include
some fraction of the galaxies' light and will doubly compound this
error.  In addition, because galaxy apertures start to significantly
overlap in our images and the HDF images when they extend to $r \sim 4
r_{\rm iso}$, we find that some flux from the wings of {\it detected}
galaxies will inevitably contribute a pedestal level to the mean sky
level in any image.  We have estimated this pedestal level by Monte
Carlo simulations as described in Appendix B of Paper I.  The pedestal
is independent of field, but does dependent on the detection limits
and surface brightness characteristics of the data. For $V_{606}$ HDF
images, this pedestal level is roughly 10\% of the total flux from
$V>23$ AB mag galaxies and, again, this error is compounded by the
fact that any flux at radii beyond galaxy apertures can be include in
the ``sky'' estimate.  The true flux from $V>23$ AB mag galaxies in
the HDF is therefore almost twice what is recovered by standard
methods (see \S \ref{numcnts}).

Using different values of $V_{\rm cut}$, we can quantify systematic
errors in faint galaxy photometry as a function of the isophotal
surface brightness limit of the data, $\mu_{\rm iso}$, and the central
surface brightness of the source, $\mu_0$.  The smaller the value of
$\Delta\mu = \mu_{\rm iso} - \mu_0 $ is for a particular galaxy, the
larger the photometric error in standard aperture photometry.  This
problem has been discussed in the literature at length with respect to
low surface brightness galaxies at low redshifts based on
extrapolations of simple exponential light profiles (Disney 1976,
Disney \& Phillips 1983, Davies 1990); the same principle begins to
apply to normal surface brightness galaxies which have low {\it
apparent} surface brightness at higher redshifts due to $(1+z)^4$
surface brightness dimming (Dalcanton 1998).

Finally, we note that the direct measurements of the EBL23 in our data
--- based on surface photometry of the total integrated background,
zodiacal light, and diffuse galactic light --- are
$2\sigma-3\sigma$ detections.
However, the mean flux from detected sources is obviously an absolute
minimum value for the EBL.  
Therefore, the strongest lower limit we can place on the flux from sources
fainter than $V=23$ AB mag (EBL23) is  the mean flux in {\it
detected} galaxies as measured by ensemble aperture photometry in our
WFPC2 data and the HDF.
The strongest upper limits we can place on EBL23 are the $2\sigma$
upper limits of our direct measurements of the EBL23.
In Table \ref{tab:cum.errors} we list 
(1) our direct measurements of the EBL23, and (2) the minimum
values for the EBL23 (minEBL23) as identified by ensemble aperture
photometry of detected sources in our WFPC2 data and the HDF.  For
comparison, the flux in published HDF galaxy counts and ground based
counts are also listed there.

\section{Comparison with Number Counts and Luminosity Functions}\label{opt.res}

Whether the light originates from stellar nucleosynthesis, accretion
onto compact objects, or gravitationally collapsing stellar systems,
the total optical flux escaping from detected galaxies is quantified
by number counts and luminosity functions.  To the detection limits,
number counts and luminosity functions contain exactly the same
information regarding the integrated background light: the integrated
flux from resolved sources is the same whether or not you know the
redshift of the sources.  However, in the context of predicting the
EBL flux, luminosity functions contain information about the intrinsic
flux distribution of the sources and thus allow us to estimate the
flux from sources beyond the detection limits with better defined
assumptions.  In the following sections, we compare our EBL detections
with the integrated flux obtained by both methods.  Dust obscuration
in the emitting sources will clearly reduce the UV and optical flux
which escapes, but the EBL, number counts, and luminosity functions
are all measurements of the {\it escaping} flux; the relative
comparisons discussed in this section are therefore independent of
dust extinction.

\subsection{Number Counts}\label{numcnts}

\begin{figure}[t]
\begin{center}
\includegraphics[width=3in,angle=0]{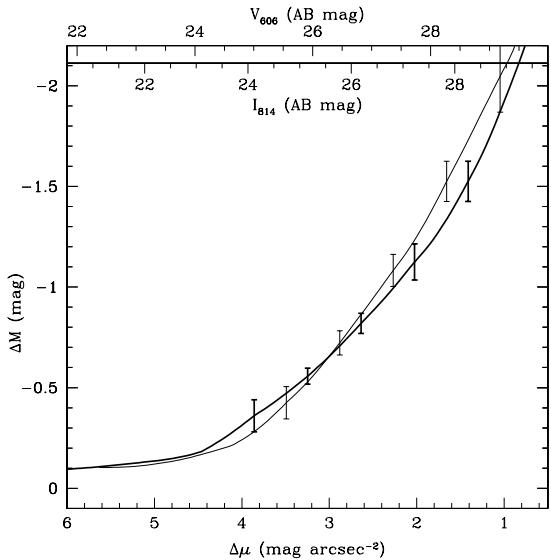}
\caption{\footnotesize
Aperture corrections as a function of $\Delta\mu = \mu_{\rm iso} - \mu_0$
(isophotal minus central surface brightness) derived by ``ensemble
aperture photometry'' of the EBL field for $V$ (thick line) and $I$
(thin line). 
The mean apparent magnitude in $V_{606}$ and $I_{814}$
corresponding to a particular value of $\Delta\mu$ 
in the HDF images is indicated by the 
x-axes at the top of the plot.
Error bars show the $1\sigma$ statistical error
in the mean corrections derived from 18 WFPC2 images of the EBL
field (six exposures, three WF chips).}
\label{fig:corfac}
\end{center}
\end{figure}

\begin{figure}[t]
\begin{center}
\includegraphics[width=3in,angle=0]{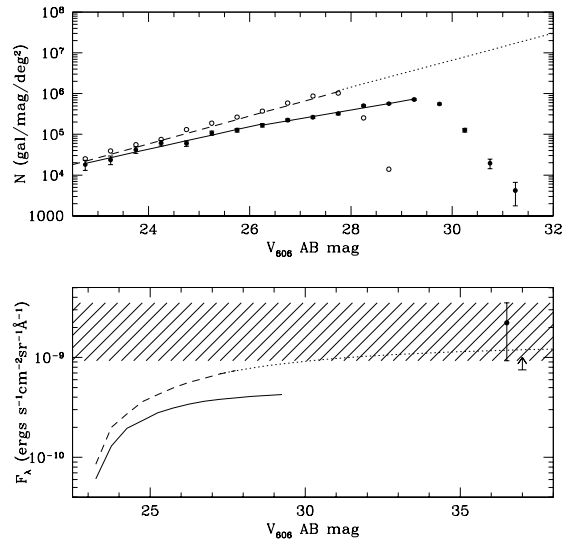}
\caption{\footnotesize The upper panel shows the galaxy counts as
published in the HDF catalog (filled circles) with $\sqrt{N}$ error
bars and the corrected number counts (open circles), as described in
\S\ref{numcnts}. The solid lines show fits to the raw number counts,
which change slope slightly around $V_{606}=26$.  The fit to the
corrected counts is indicated by the dashed line to the detection
limit and a dotted line beyond.  No change in slope is apparent at the
faint end for the corrected counts.  All slopes are given in the text.
In the lower panel, we plot the integrated flux corresponding to the
galaxy counts with the same line types as in the upper panel.  The
data point and $1\sigma$ error bar mark the value of EBL23 (converted
to $V_{606}$ from the $V_{555}$ band). The corresponding $\pm 1\sigma$
error range is emphasized by the hatch-marked region.  For comparison,
the lower limit arrow shows $2\sigma$ lower limit of minEBL23, the
integrated flux from detected sources with $V_{606}>23$ AB mag.}
\label{fig:numcntsV}
\end{center}
\end{figure}

\begin{figure}[t]
\begin{center}
\includegraphics[width=3in,angle=0]{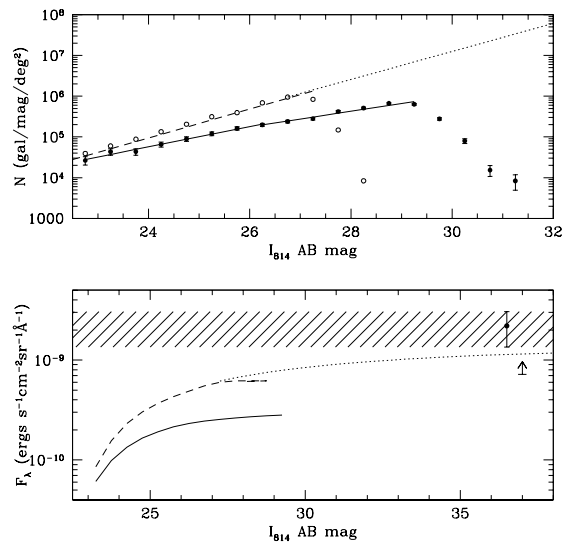}
\caption{\footnotesize The same as Figure \ref{fig:numcntsV}, but for
the $I_{814}$--band.  As for the $V_{606}$--band, the raw $I_{814}$
counts show a slight change in slope around 24--26 AB mag, while the
corrected counts do not.  All slopes are given in the text. The
integrated flux of the raw and corrected counts are compared to our
EBL23 $I_{814}$ detections in the lower panel, as in Figure
\ref{fig:numcntsV}.}
\label{fig:numcntsI}
\end{center}
\end{figure}

Using ``ensemble aperture photometry'' to measure the total flux from
galaxies as a function of magnitude in our $V_{555}$ and $I_{814}$
images of the EBL field, we find that the standard photometry
methods used to produce the HDF catalog systematically underestimate
the flux from each source, as summarized in \S\ref{sum.detect} (see
\S10 and Appendix B of Paper I for a thorough discussion).  We use
these results to derive flux corrections as a function of $\Delta\mu =
\mu_{\rm iso} - \mu_0 $ (isophotal minus central surface brightness),
which are essentially aperture corrections.  These aperture
corrections are similar to those found by other authors (c.f. Smail
\etal 1995) and are a natural consequence of integrating an extended
light profile to an insufficient radius.  This effect can be
quantified for exponential or de Vaucouleurs profiles, as in Dalcanton
(1998).  However, the corrections we show here are empirical
measurements and assume nothing about the light profiles of the
sources.   

The corrections we derive for the two bandpasses (see Figure
\ref{fig:corfac}) are very similar functions of $\Delta\mu$, which
indicates that the profiles of detected galaxies are not a strong
function of wavelength over the baseline of observed $V$ to $I$.
However, we note that a particular value of $\Delta\mu$ occurs at a
brighter AB magnitude in $I_{814}$ than in $V_{606}$ because the
limiting isophotal level (sky noise) in $I_{814}$ is 0.6 AB mag
brighter than in $V_{606}$.  The corrections are therefore larger in
$I_{814}$ than they are at the same AB magnitude at $V_{606}$.  The
corrections in both bands include a correction which compensates for
overestimates in the sky flux from foreground sources (the pedestal
sky level described in \S\ref{sum.detect}).  This correction, which
accounts for errors in the local sky estimate, ranges from  0.1--0.3
mag, monatonically increasing towards fainter magnitudes. As in
$V_{606}$ and $I_{814}$, aperture corrections for $U_{300}$ band will
depend on the profiles of galaxies at $U_{300}$ and the surface
brightness limits of the data. However, the very low signal--to--noise
ratio of our F300W images prevents us from determining aperture
corrections in that bandpass.  The $U_{300}$ photometry is discussed
further below.

We have applied the aperture corrections we derive to the individual
objects in the HDF $V_{606}$ and $I_{814}$ catalogs (Williams \etal\
1996), which fractionally increases the flux of each galaxy.  For
example, while galaxies in the HDF catalog with $V_{606}\sim30$
\ABmag\ have well-detected cores, less than 30\% of their total flux
is recovered: the total flux of a galaxy measured to have
$V_{606}\sim30$ \ABmag\ by standard photometry methods is actually
closer to $V_{606}\sim28$ \ABmag. The corrected and uncorrected (raw)
galaxy counts and corresponding integrated flux with magnitude are
compared in Figures \ref{fig:numcntsV} and \ref{fig:numcntsI}.  The
integrated flux of the corrected galaxy counts roughly corresponds to
the minimum value of EBL23, as the aperture corrections were derived
from the calculation of the minimum EBL23 in our data. Statistical
variations in galaxy counts between fields are to be expected.

The aperture corrections we apply clearly have a significant impact on
the slope of faint number counts.  To quantify this, we fit both the
raw and corrected number counts with the usual relationship between
apparent magnitude and surface number density, $N \propto 10^{\alpha
m}$, where $N$ is the number of galaxies per magnitude per square
degree.  For the raw $V_{606}$ counts, we find that the data exhibit a
change in slope around 24--26 AB mag.  A single fit
over the range $22<V_{606}<29.5$ AB mag gives $\alpha =0.24 \pm 0.01$
with a $\chi^2$ per degree of freedom ($\chi^2/{\rm dof}$) of $1.5$.
Fitting the counts brighter and fainter than 26 AB mag,
respectively, we find $\alpha_{b}=0.28 \pm 0.02$ with $\chi^2/{\rm
dof} = 0.9$ and $\alpha_{f}=0.21 \pm 0.01$ with $\chi^2/{\rm dof}=1.2$
(solid lines in the upper panel of Figure \ref{fig:numcntsV}).  We
ascribe this change in slope to the onset of significant photometry
errors.

For the corrected  $V_{606}$, counts we find that the full
$22<V_{606}<27.5$ AB  mag range is well fit by a slope of $\alpha=0.33
\pm0.01$ with $\chi^2/{\rm dof} = 0.60$ (dashed line in the upper
panel of Figure \ref{fig:numcntsV}).  This result suggests that
photometry errors are responsible for the change in slope at the
faint end of the HDF counts,
and that $N(m)$ does not, in fact, significantly decline before the
detection limit of the HDF at $V_{606}$.  In addition, while the
integrated flux in the raw galaxy counts has converged by the apparent
detection limit of the HDF, the flux from the {\it corrected} galaxy counts
has not (see the lower panel of Figure \ref{fig:numcntsV}).

\begin{figure}[t]
\begin{center}
\includegraphics[width=3in,angle=0]{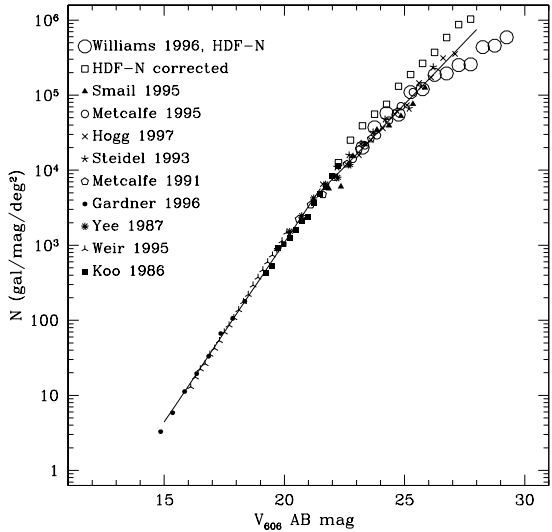}
\caption{\footnotesize
The raw and corrected number counts from Figure \ref{fig:numcntsV}
compared to number counts from the literature, labeled by first
author.  The lines indicate fits to the data using the relation $N\propto
10^{\alpha m}$; $\alpha = 0.48\pm0.1$ and $\alpha = 0.33\pm0.01$ at
the bright and faint ends, respectively. Note that the slope
of the corrected HDF
counts is well matched to that at brighter magnitudes.}
\label{fig:bright.cntsV}
\end{center}
\end{figure}

We find similar results for the $I_{814}$ counts (see Figure
\ref{fig:numcntsI}). As for $V_{606}$, the raw $I_{814}$ counts
display a slight change in slope around 24--26 AB mag. We find a slope
of  
$\alpha_b=0.25\pm 0.01$ with $\chi^2/{\rm dof} = 0.6$, and 
$\alpha_f=0.19\pm 0.02$ with $\chi^2/{\rm dof} = 2.0$,
brighter and fainter than 26 AB mag, respectively.
For the full range $22<I_{814}<29.5$ AB mag, we find 
$\alpha=0.22\pm 0.01$ with $\chi^2/{\rm dof} = 2.1$.
For the corrected  $I_{814}$ counts, we find
$\alpha=0.34\pm 0.01$ with $\chi^2/{\rm dof} = 0.8$ at $22<I_{814}<27$
AB mag.  

In Figure \ref{fig:bright.cntsV}, we show the raw and corrected HDF
counts relative to $V$- and $R$-band counts available in the
literature for $V>15$ AB mag.  We have converted all of the published
counts to $V$-band AB mag by applying constant offsets consistent with
those in Fukugita, Shimasaku, \& Ichikawa (1995). These incorporate mean
K--corrections based on the mean redshift corresponding to the
apparent magnitude of the sample.  Differences between filters will
have some affect on the slope of counts in surveys which cover a large
range of redshift (apparent magnitude) due to changing galaxy colors
and K-corrections with increasing redshift, but these effects will
average out between the multiple surveys shown. This plot shows that
the aperture corrections we have applied to the HDF sources produce
number counts which have a slope consistent with the slope found at
brighter magnitudes.  

\begin{figure}[t]
\begin{center}
\includegraphics[width=3in,angle=0]{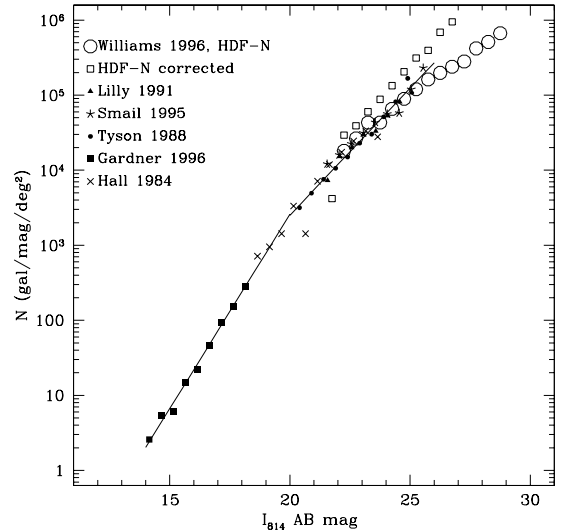}
\caption{\footnotesize
Same as Figure \ref{fig:bright.cntsV} but for the $I-$band 
counts. The lines indicate fits to the data using the relation $N\propto
10^{\alpha m}$; $\alpha = 0.52\pm0.1$ and $\alpha = 0.34\pm0.01$ at
the bright and faint ends, respectively. Note that slope of the
corrected HDF counts is well matched to the slope at brighter
magnitudes, and that the slope of the $I$ and $V$ band counts are similar
at all magnitudes.}
\label{fig:bright.cntsI}
\end{center}
\end{figure}

In Figure \ref{fig:bright.cntsI}, we show the same plot for the
$I$--band.  Again, the corrected $I_{814}$ counts display a slope
which is similar to that found at magnitudes brighter than 23 AB mag.
Note also that slope of the counts at $<25$ AB mag in $V$ and $I$ are
the same to within the statistical errors.  The aperture corrections
we apply to the HDF counts at $V_{606}$ and $I_{814}$ extend this
agreement to the current detection limits.  The corrected counts imply
that the faintest galaxies detected do not exhibit a significantly
steeper slope in $V_{606}$ than in $I_{814}$, in contrast with the raw
galaxy counts. This is an important constraint on galaxy evolution
models.

Although the signal to noise in the $U_{300}$ data is too low to allow
us to obtain accurate aperture corrections at that wavelength, the
minimum EBL23 at $U_{300}$ implies consistent colors for faint and
bright galaxies at $U-V$, as in $V-I$ (see Figure \ref{fig:eblsummary}
and Table \ref{tab:cum.errors}).  We note, also, that the color of the
integrated flux from galaxies is consistent with the color of the
total background light within $2\sigma$.  In other words, no exotic
population of sources is required to produce the detected background.

The lack of turnover in the corrected counts strongly suggests that
sources do exist at apparent magnitudes beyond the present detection
limit.  If we impose no limit on the apparent magnitude of sources and
simply extrapolate the galaxy counts beyond $V_{606}\sim27.5$ AB mag
using $\alpha=0.33$ (dotted line in Figure \ref{fig:numcntsV}), we
obtain a prediction for the total integrated EBL23 of $1.3$\tto{-9}
cgs, which is $1\sigma$ below the measured value in the EBL field.
In this case, the predicted EBL23 converges around $V_{606}\sim50$ AB
mag, which is significantly fainter than a dwarf galaxy at $z\sim
6$. However, very little flux is obtained from the faintest bins.  If
we impose the limit $V_{606}\sim 38$ AB mag as the faintest apparent
magnitude for a realistic source (e.g., a dwarf galaxy with $M_{V}\sim
-10$ AB mag at $z\sim4$), we obtain a flux of 1.2\tto{-9} cgs.
The flux from sources with $I_{814}>23$ AB mag is $1.3$\tto{-9} cgs if
we adopt $\alpha = 0.34$, with the flux converging around $I_{814}\sim
60$ AB mag.  Adopting a more realistic faint cut--off of $\sim 38$ AB
mag, as discussed for $V_{606}$, we obtain a total flux of
$1.2$\tto{-9} cgs, $1\sigma$ below the mean detected value of
EBL23 at $I_{814}$ (see Figure \ref{fig:numcntsI}).

In order to obtain a cumulative flux equal to the mean detected EBL
(or the upper limit) from sources brighter than $\sim 38$ mag, the
slope of the galaxy counts in the range 28--38 AB mag would clearly
need to {\it increase} at some point beyond the current detection
limit.  
For example, the total flux from sources $23<V_{606}<38$ AB mag will
produce the mean detected EBL if the sources with $23<V_{606}<28$ AB
mag obey a slope of $\alpha = 0.33$ and sources with $28<V_{606}<38$
AB mag obey $\alpha = 0.42$.  We stress, however, that the total flux
obtained from sources with such a steep faint--end slope is critically
dependent on the cut of magnitude: the total flux reaches
$5.1$\tto{-9} cgs if we integrate the counts to $50$ AB mag, and to
9.0\tto{-9} cgs if we integrate to $60$ AB mag.
Recall that our $2\sigma$ upper limit on EBL23 at $V_{606}$ is
$5.0$\tto{-9} cgs.  For $\alpha=0.35$ at $V>28$ AB mag, the integrated
flux reaches 1.37, 1.51, and 1.57\tto{-9} cgs (converged) for faint
cut--off magnitudes of 40, 60, and 80 AB mag respectively.  Although
it is obviously impossible to place firm constraints on the number
counts beyond the detection limit, as they may change slope at any
magnitude, we conclude that it is very unlikely that the slope beyond
$V_{606}\sim28$ AB mag is steeper than $\alpha = 0.42$.  If the slope
continues at $0.33<\alpha< 0.35$, the EBL23 reaches a roughly
1.3--1.5\tto{-9} cgs by $V_{606}\sim40$ AB mag, $<1\sigma$ below our
detected value.

Similarly, for the $I$--band the integrated flux from sources matches
the mean detected EBL23 if the sources with $23<I_{814}<27$ AB mag
obey a slope of $\alpha = 0.34$ and sources with $27<I_{814}<39$ AB
mag obey $\alpha = 0.42$. For those slopes, the total flux reaches the $2\sigma$ upper
limit of the EBL23 at $I_{814}$ by 50 AB mag.  For
$\alpha=0.36$, slightly above the slope we find for the corrected
counts, the integrated flux reaches 1.31, 1.58, and 1.62\tto{-9} cgs 
(converged) for faint cut--off magnitudes of 40, 60, and 80 AB mag, 
respectively.  As for the $V$--band, we conclude that it is unlikely
that the $I$--band faint--end slope is steeper than 0.42 at any magnitude.
For a slope of $0.34< \alpha< 0.36$  for $I>27$ AB mag, the
EBL reaches 1.2--1.3\tto{-9} by $I_{814}\sim 40$ AB mag, $1\sigma$ below our
detected value.

In summary, we conclude from the corrected number counts shown in
Figures \ref{fig:numcntsV} -- \ref{fig:bright.cntsI} that sources are
likely to exist beyond the detection limit of the HDF.  Furthermore,
if the number counts continue with the slope we measure at the
faintest levels, then the predicted EBL23 is within $1\sigma$ of the
detected EBL23 at both $V_{606}$ and $I_{814}$.  If we extrapolate
beyond the detection limits assuming the slope found from the
corrected number counts, we find that less than 50\% of EBL23 comes
from sources beyond the current detection limit at $V_{606}$ or
$I_{814}$ --- the majority of the light contributing to EBL23 comes
from sources which can be individually detected.

Finally, we note that our ensemble photometry method yields a
statistical correction for the light lost from the wings of galaxies
beyond the detection isophote.  This light cannot, by definition, be
recovered by standard single-object photometry.  In contrast, the
ensemble photometry method effectively adds together the light beyond
the detection isophote from many galaxies to enable a significant
measurement of that light. 

\subsection{Luminosity Functions}\label{lumfuncs}

\begin{figure}[t]
\begin{center}
\includegraphics[width=3in,angle=0]{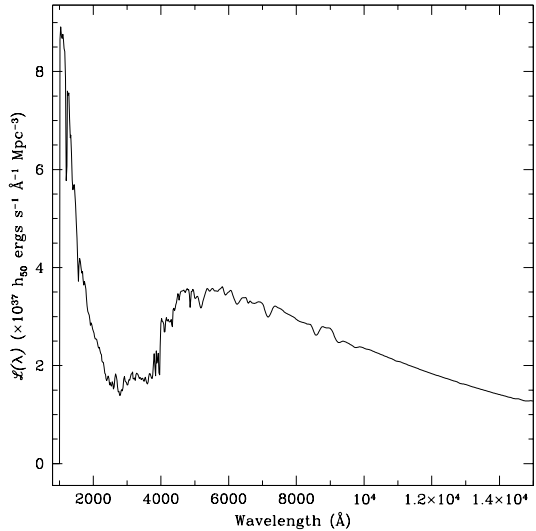}
\caption{\footnotesize
The local luminosity density constructed by combining the spectral
energy distributions of E/S0, Sab, and Sc galaxies weighted according
to the type-dependent luminosity functions as described in
\S\ref{lumfuncs} and Equation \ref{eq:local.lumden}.}
\label{fig:locallumden}
\end{center}
\end{figure}

\begin{figure}[t]
\begin{center}
\includegraphics[width=3in,angle=0]{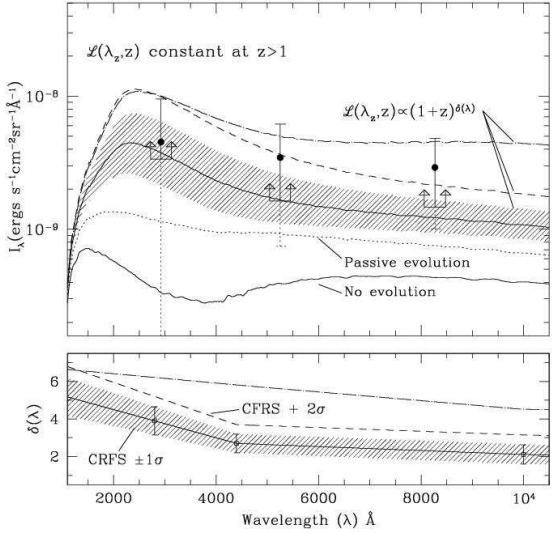}
\caption{\footnotesize
The upper panel shows the spectrum of the EBL calculated by
integrating the luminosity density over redshift (Equation
\ref{eq:vol.int}) for constant luminosity density, passively evolving
luminosity density, and evolution of the form ${\cal L}(\lambda,z) =
{\cal L}(\lambda,0) (1+z)^{\delta(\lambda)}$, with ${\cal
L}(\lambda,z)$ constant at $z>1$. The lower panel shows
$\delta(\lambda)$ for the three cases of $(1+z)^{\delta(\lambda)}$ as
labeled in the figure and described in the text.  Line types and
hatch-marked regions in upper panel correspond to values of
$\delta(\lambda)$ with the same line type in the lower panel.
Filled circles show the mean EBL detections with $2\sigma$ error bars.
The error bars are dashed where they extend below the cumulative
flux in detected sources --- the minimum values for the
EBL --- indicated by the lower limit brackets.}
\label{fig:lumden}
\end{center}
\end{figure}

\begin{figure}[t]
\begin{center}
\includegraphics[width=3in,angle=0]{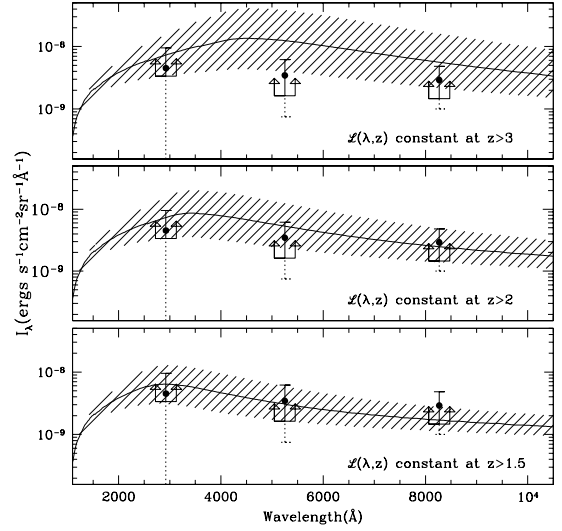}
\caption{\footnotesize
The three panels show the spectrum of the EBL calculated assuming
${\cal L}(\lambda,z) = {\cal L}(\lambda,0) (1+z)^{\delta(\lambda)}$
over the range $0<z<1.5$ (bottom panel), $0<z<2$ (middle), and $0<z<3$
(top). In all cases, the luminosity density is held constant beyond
the indicated redshift limit.  The hatch-marked regions each show the
$\pm1\sigma$ range of CFRS values for
$\delta(\lambda)$, as in Figure \ref{fig:lumden}. The integrated EBL
as a function of redshift is shown in Figure \ref{fig:eblwithz}.
Luminosity density as a function of redshift is shown in Figure
\ref{fig:lumdenwithz} for some combinations of $\delta(\lambda)$ and
the redshift cut-off for evolution.  The filled circles, error bars
and lower limit symbols are as in Figure \ref{fig:lumden}.}
\label{fig:lumdenbeyond1}
\end{center}
\end{figure}

\begin{figure}[t]
\begin{center}
\includegraphics[width=3in,angle=0]{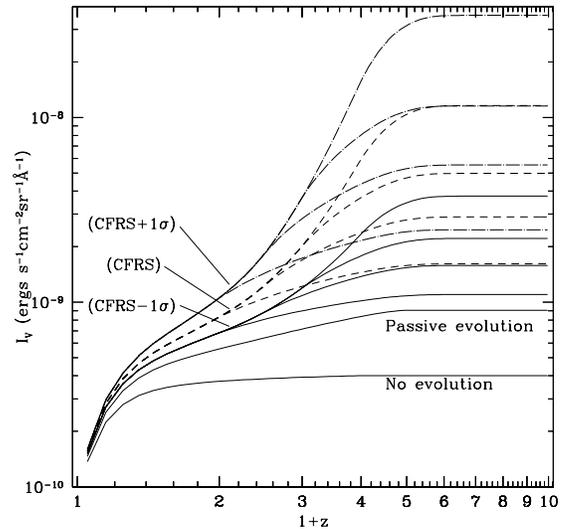}
\caption{\footnotesize
The integrated EBL at $V_{555}$ contributed as a function of
increasing redshift from $z=0$ to $z=10$. As marked in the figure, 
the lines show the
integrated flux for no evolution in the luminosity density,
passive evolution, and evolution of the form 
${\cal L}(\lambda,z) \propto
(1+z)^{\delta(\lambda)}$ for the $-1\sigma$, mean, 
and $+1\sigma$ range of $\delta(\lambda)$
values determined by CFRS. For each
$\delta(\lambda)$, we plot the  growth of the  EBL with redshift 
if ${\cal L}(\lambda,z)$ is held constant at $z>1$, $z>1.5$, $z>2$, and $z>3$,
corresponding to Figures \ref{fig:lumden} and \ref{fig:lumdenbeyond1}.}
\label{fig:eblwithz}
\end{center}
\end{figure}

In this section, we compare the detected EBL with the EBL predicted by
luminosity functions measured as a function of redshift.  To avoid
unnecessary complications in defining apparent magnitude cut-offs, and
to facilitate comparison with other models of the luminosity density
as a function of redshift, we compare luminosity functions with the
total EBL rather than with EBL23, as in the previous section.  To do
so, we combine the EBL23 flux measured in Paper I with the flux from
number counts at brighter magnitudes, as given in Table
\ref{tab:cum.errors}.
Systematic errors in photometry of the sort discussed in
\S\ref{numcnts} are likely to be relatively small for redshift surveys
because the objects selected for spectroscopic surveys are much
brighter than the limits of the photometric catalogs (although see
Dalcanton 1998 for discussion of the effects of small, systematic
photometry errors on inferred luminosity functions). We have not
tried to compensate for such effects here.

The integrated flux from galaxies at all redshifts is given by 
\begin{equation}\label{eq:vol.int}
 I(\lambda,0) = \int_{0}^{z}{
	{\cal L}(\lambda_z,z) { dV_c(z) \over 4\pi D_L(z)^2 } }
\end{equation}
in which $V_c(z)$ is the comoving volume element, $D_L(z)$ is the
luminosity distance, $\lambda_0$ is the observed wavelength, and
$\lambda_z = \lambda_0 (1+z)^{-1}$ is the rest--frame wavelength at
the redshift of emission.  To compare the detected EBL to the observed
luminosity density with redshift, ${\cal L}(\lambda,z)$, we begin by
constructing the SED of the {\it local} luminosity density as a linear
combination of SEDs for E/S0, Sb, and Ir galaxies, weighted by their
fractional contribution to the local $B$--band luminosity density:
\begin{equation}\label{eq:local.lumden}
{\cal L}(\lambda, 0) = \sum_{i}{ {\cal L}_i(B,0) 
				 {f_i({\lambda}) \over f_i(B)}
				},
\label{eq:make.lumden}
\end{equation}
in which the subscript $i$ denotes the galaxy Hubble type (E/S0, Sb,
or Ir), $f_i(\lambda)$ denotes the galaxy SED (the flux per unit
rest--frame wavelength), and ${\cal L}_i(B,0)$ is the $B$-band, local
luminosity density in ergs s$^{-1}$ \AA$^{-1}$ Mpc$^{-3}$.  To produce
the integrated spectrum of the local galaxy population, we use
Hubble--type--dependent luminosity functions from Marzke \etal (1998)
and SEDs for E, Sab, and Sc galaxies from Poggianti (1997).  We adopt
a local luminosity density of ${\cal L}_B = 1.3\times10^8 h
L_{\odot}$Mpc$^{-3}$, consistent with the Loveday \etal (1992) value
adopted by CFRS and also with Marzke \etal (1998).\footnote{For a
$B$-band solar irradiance of $L_\odot=4.8\times10^{29}$\esa, ${\cal
L}(B,0) = 6.1\times10^{37} h$ ergs s$^{-1}$ Mpc$^{-3} =
4.0\times10^{19} h_{50}$ W Hz$^{-1}$ Mpc$^{-3}$.}  The spectrum we
obtain for ${\cal L}(\lambda,0)$ is shown in Figure
\ref{fig:locallumden}.  

We note that the recent measurement of the local luminosity function
by Blanton \etal (2001) indicates a factor of two higher local
luminosity density than found by previous authors. Previous results
are generally consistent with Loveday \etal to within 40\%.  Blanton \etal
attribute this increase to deeper photometry which recovers more flux
from the low surface brightness wings of galaxies in their sample
relative to previous surveys (see discussions in \S\ref{numcnts}), and
photometry which is unbiased as a function of redshift.  For the
no-evolution and passive evolution models discussed below, the
implications of the Blanton \etal results can be estimated by simply
scaling the resulting EBL by the increase in the local luminosity
density.  Although the Blanton \etal results do not directly pertain
to the luminosity functions measured by CFRS at redshifts $z>0.2$,
they do suggest that redshift surveys at high redshifts will
underestimate the luminosity density, as discussed by Dalcanton
(1998).

In the upper panel of Figure \ref{fig:lumden}, we compare the EBL flux
we detect with EBL flux predicted by five different models for ${\cal
L}({\lambda}, z)$, using the local luminosity density derived in
Equation \ref{eq:local.lumden} as a starting point.  
For illustrative purposes, the first model we plot shows the EBL which
results if we assume no evolution in the luminosity density with
redshift, i.e. ${\cal L}({\lambda}, z) = {\cal L}(\lambda, 0)$.  The
number counts themselves rule out a non-evolving luminosity density,
as has been discussed in the literature for over a decade;
inconsistency between the detected EBL and the no-evolution model is
just as pronounced.  The predicted EBL for the no-evolution model is a
factor of 10 fainter than the detected values (filled circles).  These
are $1.7\sigma$, $2.1\sigma$, and $2.2\sigma$ discrepancies at
$U_{300}$, $V_{555}$ and $I_{814}$, respectively. More concretely, the
no-evolution prediction is at least a factor of $12\times$, $4\times$,
and $3.7\times$ lower than the flux in {\it individually resolved}
sources at $U_{300}$, $V_{555}$ and $I_{814}$ (lower--limit arrows).
Note that the no-evolution model still underpredicts the EBL if we
rescale the local luminosity density to the Blanton \etal (2001)
values. This model demonstrates the well-known fact that luminosity
density is larger at higher redshifts.

The second model we plot in Figure \ref{fig:lumden} shows the effect
of passive evolution on the color of the predicted EBL.  In this
model, we have used the Poggianti (1997) SEDs for galaxies as a
function of age for $H_0= 50$ km s$^{-1}$ Mpc$^{-1}$ and $q_0 =
0.225$.  In the Poggianti models, stellar populations are 2.2 Gyrs old
at a $z\sim3$.  The resulting ${\cal L}({\lambda}, z)$ is bluer than
the no-evolution model due to a combination of K-corrections and
increased UV flux for younger stellar populations. The passive
evolution model does provide a better qualitatively match to the SED
of the resolved sources (lower limits) and EBL detections (filled
circles); however, it is still a factor of $3\times$ less than the
flux at $U_{300}$, and a factor of $2\times$ less than the flux we
recover from resolved sources at $V_{555}$ and $I_{814}$.  For the
adopted local luminosity density and Poggianti models, passive
evolution is therefore not sufficient to produce the detected
EBL. Again, the passive evolution adopted here still underpredicts the
EBL if we rescale the local luminosity density by a factor of two to
agree with the Blanton \etal (2001) value.

As a fiducial model of evolving luminosity density, we adopt the form
of evolution implied by the CFRS redshift survey (Lilly \etal 1996,
hereafter CFRS) and Lyman-limit surveys of Steidel \etal (1999):
${\cal L}({\lambda},z) ={\cal L}(\lambda,0) (1+z)^{\delta(\lambda)}$
over the range $0<z<1$ and roughly constant luminosity density at
$1<z<4$.  The remaining three models shown in Figure \ref{fig:lumden}
test the strength of evolution of that form which is allowed by the
EBL detections. The hatch-marked region shows the EBL predicted for
values of $\delta(\lambda)$ which represent the $\pm 1\sigma$ range
found by CFRS for the redshift range $0<z<1$.  The value of the
exponent $\delta(\lambda)$ is indicated in the lower panel of Figure
\ref{fig:lumden}, and the hatch--marked region reflects the
uncertainty in the high redshift luminosity density due to the poorly
constrained faint--end slope of the luminosity functions.  This $\pm
1\sigma$ range of the predicted EBL is consistent with the detected
EBL at $U_{300}$, but is inconsistent with the EBL detections at
$V_{555}$ and $I_{814}$ at the $1\sigma$ level of both model and
detections. It is, however, consistent with the integrated flux in
detected sources at $V_{555}$ and $I_{814}$.

\begin{figure}[t]
\begin{center}
\includegraphics[width=3in,angle=0]{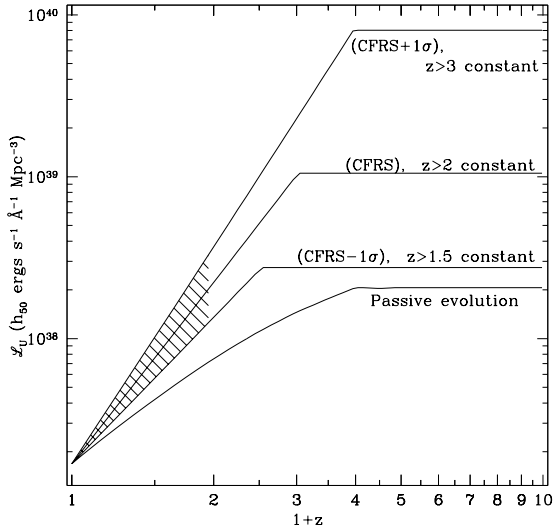}
\caption{\footnotesize
The luminosity density at $U_{300}$ as a function of redshift
corresponding to limiting cases plotted in Figure \ref{fig:lumdenbeyond1}.
The hatch--marked region indicates the $\pm 1\sigma$ range given by
the CFRS measurements of $\delta(\lambda)$ over the range $0<z<1$.
The horizontal line segments show the luminosity density corresponding
to the $-1\sigma$ limit for $\delta(\lambda)$ held constant at $z>1.5$;
the mean value for $\delta(\lambda)$ held 
constant at $z>2$; and the $+1\sigma$ limit
for $\delta(\lambda)$ held constant at $z>3$.}
\label{fig:lumdenwithz}
\end{center}
\end{figure}

To test the range of evolution allowed by the full $\pm2\sigma$ range
of the EBL detections, we can explore two possibilities:  (1) stronger evolution at
$0<z<1$, shown in Figure \ref{fig:lumden}; and (2) evolution continuing beyond
$z=1$, shown in Figures \ref{fig:lumdenbeyond1}, \ref{fig:eblwithz}, and
\ref{fig:lumdenwithz}.  Addressing the possibility of
constant luminosity density at $z>1$, the
dashed line in the upper panel of Figure \ref{fig:lumden} shows the
EBL predicted by the $2\sigma$ upper limit for $\delta(\lambda)$ from
CFRS; the dot--dashed line corresponds to the value of
$\delta(\lambda)$ required to obtain the upper limits of the EBL
detections at all wavelengths.  Note that the latter implies a value
for ${\cal L}(4400\AA,1)$ which is $\sim10\times$ higher than the
value estimated by CFRS.  This result emphasizes that the $2\sigma$
interval of the EBL detections span a factor of $4$ in flux at
4400\AA, and thus the allowed range in the luminosity density for
$\lambda<4400$\AA\ and $0<z<1$ is similarly large.  Also, for each
model in which the luminosity density is constant at $z>1$, less than
$50$\% of the EBL will come from beyond $z=1$ due to the combined
effects of K-corrections and the decreasing volume element with
increasing redshift (see Figure \ref{fig:eblwithz}).

Evolution continuing beyond $z=1$ is possible if the
Lyman--limit-selected surveys have not identified all of the star
formation at high redshifts, and estimates of the luminosity density
at $3\lta z \lta 4$ are subsequently low.  Figures
\ref{fig:lumdenbeyond1} and \ref{fig:eblwithz} show the EBL predicted
by models in which the luminosity density increases as
$(1+z)^{\delta(\lambda)}$ to redshifts of $z= 1.5$, 2 and 3.  Clearly,
significant flux can come from $z>1$ if the luminosity density
continues to increase as a power law.  The rest--frame $U_{300}$
luminosity density is plotted as a function of redshift in Figure
\ref{fig:lumdenwithz} for limiting values of the cut--off redshift for
evolution and $\delta(\lambda)$. Although the strongest evolution
plotted  over--predicts the detected EBL, our detections are clearly
consistent with some of the intermediate values of the
$\delta(\lambda)$ and increasing luminosity density beyond $z=1$.
For example, the mean rate of increase in the luminosity density found
by CFRS can continue to redshifts of roughly 2.5--3 without
over-predicting the EBL.

In all models, we have adopted the same cosmology ($h=0.5$ and
$\Omega=1.0$) as assumed by CFRS and Steidel \etal (1999) in
calculating ${\cal L}(\lambda,z)$ and $\delta(\lambda)$.  Although the
luminosity density inferred from these redshift surveys depends on the
adopted cosmological model, the flux per redshift interval is a
directly observed quantity. The EBL is therefore a directly observed
quantity over the redshift range of the surveys, and is also
model--independent.  To the degree that the luminosity density becomes
unconstrained by observations at higher redshifts, the EBL does depend
on the assumed (not measured) luminosity density and on the adopted
cosmology through the volume integral.  Although dependence of the
predicted EBL on $H_0$ cancels out between the luminosity density,
volume element, and distance in Equation \ref{eq:vol.int},
$H_0$ has some impact through cosmology--dependent time scales, which
affect the evolution of stellar populations.  If the luminosity
density is assumed to be constant for $z>1$, the predicted EBL
increases by 25\% at $V_{555}$ for ($\Omega_M=0.2,\Omega_\Lambda=0$)
and corresponding values of $\delta(\lambda)$, and decreases by 50\%
for (0.2,0.8).  The luminosity densities corresponding to the
$2\sigma$ upper limit of the detected EBL change by the same fractions
for the different cosmologies if $\cal L$ is constant at
$z>1$. Similarly, for models in which the luminosity density continues
to grow at $z>1$, the luminosity density required to produce the EBL
will be smaller if we adopt (0.2,0) than (1,0), and smaller still for
(0.2,0.8).  The exact ratios depend on rate of increase in the
luminosity density.

Several authors (Treyer \etal  1998, Cowie \etal  1999, and Sullivan
\etal  2000) have found that the $\cal L$ at UV wavelengths
(2000-2500\AA) is higher than claimed by CFRS (2800\AA) in the range
$0<z<0.5$ and have found weaker evolution in the UV luminosity
density, corresponding to $\delta(2000\AA)\sim 1.7$.  The implications
for the predicted EBL can be estimated from the plots of the ${\cal
L}(U_{300}, z)$ shown in Figure \ref{fig:lumdenwithz}, and the
corresponding EBL in Figures \ref{fig:lumdenbeyond1} and
\ref{fig:eblwithz}.  For instance, if the local UV luminosity density
is a factor of 5 higher than the value we have adopted 
and if $\delta(2000\AA)\sim1.7$ over the range $0<z<1$, then the
rest--frame UV luminosity density at $z=1$ is similar to that measured
by CFRS, and the predicted $U_{300}$ EBL will be roughly
3.5\tto{-9} cgs, very similar to the EBL we derive from our modeled
local luminosity density and the mean values for $\delta(\lambda)$
from CFRS.

\subsection{Discussion}\label{opt.disc}

Evolution in the luminosity density of the form
$(1+z)^{\delta(\lambda)}$ at $0<z<1$ and slower growth or stabilization
at $z>1$, such as suggested by redshift surveys at $0<z<4$, is
consistent with the detected EBL for values of $\delta(\lambda)$
consistent with CFRS.
The strongest constraints we can place on the
EBL span a factor of 5 in flux. As such, stronger evolution between
$0<z<1$ than reported by CFRS or continuing evolution at
$z>1$ cannot be tightly constrained.  We note that recent results from
Wright (2000), which constrain the 1.25\micron\ EBL flux to be
2.1($\pm1.1$)\tto{-9} cgs, are in good agreement with our results,
but do not improve the constraints on the high redshift optical
luminosity density over those discussed above.

In contrast with our results, previous authors have claimed good
agreement between the flux in the raw number counts from the HDF and
integrated flux in the measured CFRS luminosity density to $z=1$
under the assumption that the luminosity density {\it drops rapidly} at $z>1$
(Madau, Pozzetti, \& Dickinson 1998, Pozzetti \etal  1998).  In that
work, the errors in faint galaxy photometry which cause $\sim50$\%
underestimates of the total light from $V>23$ galaxies (discussed in
\S\ref{numcnts}) are compensated by the assumption that the luminosity
density drops rapidly beyond $z=1$.  That assumption was based on
measurements of the flux from Lyman--limit systems in the HDF field by
Madau et al (1996), which are substantially lower than measurements by
Steidel \etal (1999) due to underestimates of the volume corrections
and to the small--area sampling.  We find that the detected EBL is
{\it not} consistent with luminosity evolution comparable to the CFRS
measured values at $0<z<1$ if the luminosity density drops rapidly at
$z>1$.

\section{Flux from Sources Below the Surface Brightness Detection Limit}\label{opt.unres}

\begin{figure}[t]
\begin{center}
\includegraphics[width=2.5in,angle=-90]{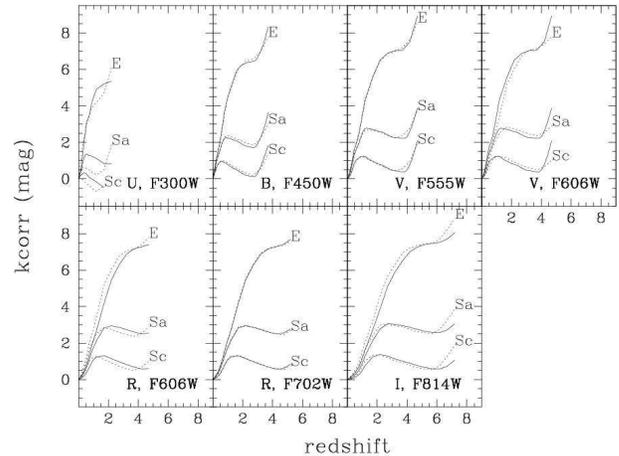}
\caption{\footnotesize
K-corrections for various filters as a function of redshift calculated
using Poggianti (1997) SEDs for present day E, Sa, and Sc
galaxies.  Bandpasses are standard Johnson/Cousins filters and the
corresponding WFPC2 filters, shown with solid and dotted lines,
respectively.}
\label{fig:k}
\end{center}
\end{figure}

The fractional EBL23 flux which comes from {\it detected} sources is
simply the ratio of the flux in detected sources (measured by
``ensemble photometry'') to the detected EBL23 ($\pm2\sigma$ limits).
The maximum fractional EBL23 flux coming from {\it undetected} sources
is what remains: 0-65\% at $U_{300}$, 0-80\% at $V_{555}$, and 0-80\%
at $I_{814}$.  Although these limits include the possibility of no
additional contribution from undetected sources, it is worthwhile to
note that if the progenitor of a normal disk galaxy at $z=0.1$
(central surface brightness $\mu_0(V) \sim 21.3$ mag arcsec$^{-2}$ and
$V\sim 22$ mag) existed at $z\sim2$, then the $z\sim 2$ progenitor
would have a core surface brightness (within $0.2\times0.2$ arcsec$^2$)
of $\sim 26$ mag arcsec$^{-2}$ for standard K--corrections
and passively evolving stellar populations (e.g. Poggianti 1997),
which is roughly the detection limit for the HDF.  In particular,
regardless of the exact evolutionary or k--corrections, dimming due to
cosmological effects alone (redshift and angular resolution) produce
$\sim5$ mag of surface brightness dimming. This effect is independent
of wavelength, so that dimming at other bandpasses is similar, modulo
differences in the evolutionary and K--corrections (shown in Figures
\ref{fig:k}, \ref{fig:e} and \ref{fig:eplusk}).  At $I$, for example,
the drop in surface brightness for a disk galaxy at $z\sim2$ relative
to $z\sim0$ is $\sim0.5$ mag greater than for $V$.  The progenitor of
a typical disk galaxy at $z=0.1$, which has $V-I\sim0.9$ and $\mu_0(I)
\sim 20.3$ mag arcsec$^{-2}$ (de Jong \& Lacey 2000), will have
$V-I\sim0.5$ at $z\sim2$ and $\mu_0(I) \sim 25.5$. Thus the typical
disk galaxy at $z\sim2$ is close to the HDF detection limit  in $I$ as
well as $V$.  Irrespective of the color evolution with redshift, the
point is that cosmological surface brightness dimming alone suggests
that a significant fraction of the EBL23 may come from normal galaxies
at redshifts $z<4$ which are undetectable in the HDF.
Furthermore, recent redshift surveys for low surface brightness (LSB)
galaxies now suggest that the distribution of galaxies in $\mu_0$ is
nearly flat for $\mu_0 > 22.0 B$ mag arcsec$^{-2}$  at
some luminosities (Sprayberry \etal
1997, Dalcanton \etal 1997, O'Neil \& Bothun 2000, Blanton \etal 2001,
Cross \etal 2001).  If such
populations exist at high redshift, they may contribute significant
flux to the EBL as presently undetectable sources.

In this section, we explore the possible contributions to the EBL23
from galaxies at all redshifts which escape detection in the HDF
because of low {\it apparent} surface brightness.  To do so, we have
simulated galaxy populations at redshifts $0<z<10$ as the passively
evolving counterparts of local galaxy populations and then
``observed'' the simulated galaxies through the Gaussian $0.1''$ FWHM
point spread function of WFPC2. We define the surface brightness
detection threshold to be consistent with the $5\sigma$ detection
limits of the HDF images (see Table \ref{tab:model.limits}), which
correspond to roughly the turn--over magnitude in the number counts.
This exercise is not meant to approximate realistic galaxy populations
at high redshift; the evolution of galaxy populations in surface
brightness, luminosity, and number density is so poorly constrained at
present that more specific modeling is unwarranted.  The models
discussed here simply address the question: how much of the total flux
from a local-type galaxy population at a redshift $z$ can be resolved
into individual sources?

\subsection{Models}

\begin{figure}[t]
\begin{center}
\includegraphics[width=2.5in,angle=-90]{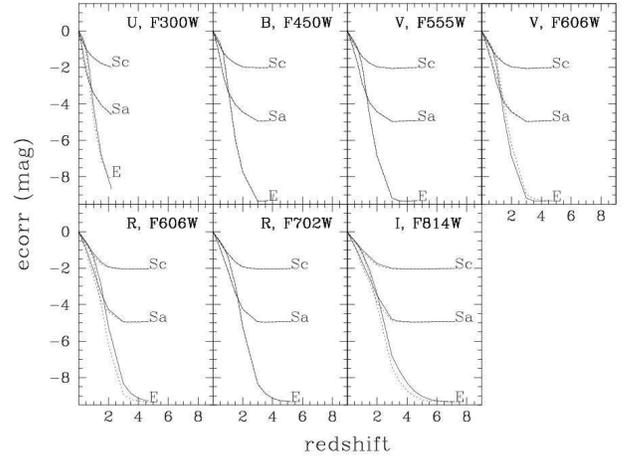}
\caption{\footnotesize
Evolutionary corrections for various filters as a function of redshift 
calculated using Poggianti (1997) SEDs for E, Sa, and Sc
galaxies evolving passively with redshift. Bandpasses are standard
Johnson/Cousins filters and the corresponding WFPC2 filters, shown
with solid and dotted lines, respectively.}
\label{fig:e}
\end{center}
\end{figure}

\begin{figure}[t]
\begin{center}
\includegraphics[width=2.5in,angle=-90]{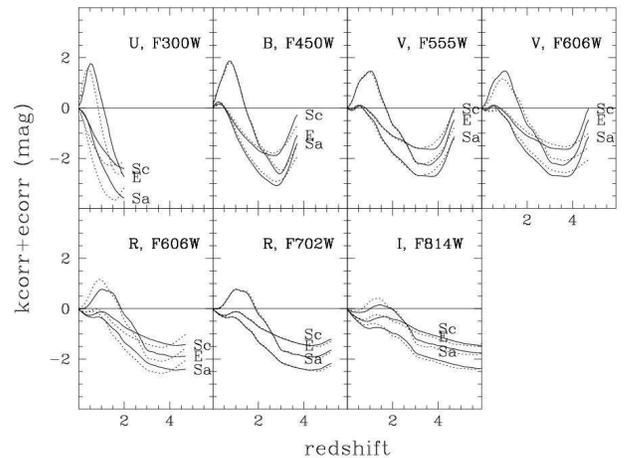}
\caption{\footnotesize The sum of the evolutionary and K--corrections
for Johnson/Cousins (solid lines) and WFPC2 filters (dotted lines)
shown in Figures \ref{fig:k} and \ref{fig:e}.}
\label{fig:eplusk}
\end{center}
\end{figure}

\begin{figure}[t]
\begin{center}
\includegraphics[width=2.5in]{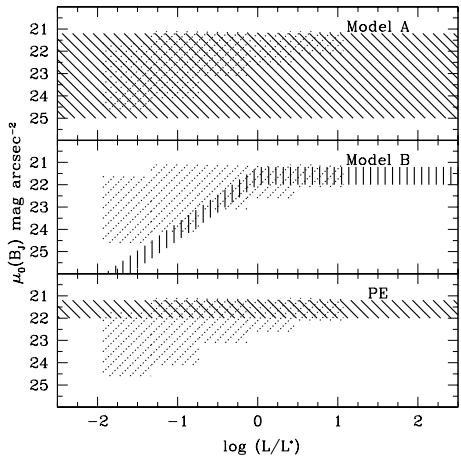}
\caption{\footnotesize The dark hatched regions in the three panels
show the surface brightness distribution as a function of luminosity
for the models adopted here (Model A, Model B, and PE).  For
comparison, the light hatched region shows the surface brightness
distribution as a function of luminosity (relative to $L^*$) as found
by Blanton \etal (2001).}
\label{fig:mods}
\end{center}
\end{figure}

\begin{figure}[t]
\begin{center}
\includegraphics[width=3in,angle=0]{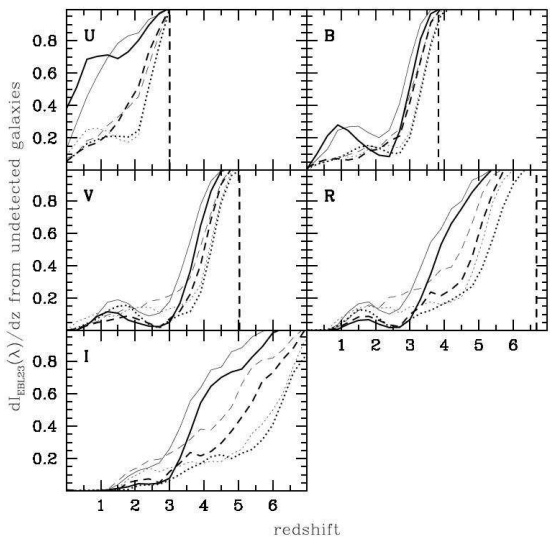}
\caption{\footnotesize
For the Johnson/Cousins bandpasses indicated in each panel, we plot
the EBL from  galaxies which are not individually detected 
in each redshift bin,
normalized by the total EBL in each redshift bin.  Models A, B and
PE are marked with solid, dashed, and dotted lines, respectively.
Thick lines correspond to simulations run with $h=0.7$,
$\Omega_0=1.0$, while thin lines correspond to $h=0.7$,
$\Omega_0=0.2$.  The dotted vertical line in each panel
indicates the Lyman limit for the band-pass.}
\label{fig:frac.det.with.z}
\end{center}
\end{figure}

We have considered three models for the central surface brightness
distributions of disk galaxies in order to explore the possible
contributions from LSB galaxies.  These models are taken from Ferguson
\& McGaugh 1995 (FM95) and can be generally described as follows: a
standard passive evolution model in which all galaxies have central surface
brightnesses in the range $21<\mu_0(B_J)<22$ mag arcsec$^{-2}$ (Model
PE); an LSB--rich model (Model A), in which galaxies of all
luminosities have $21.5<\mu_0(B_J)<25$ mag arcsec$^{-2}$; and a more
conservative LSB model (Model B), in which $\mu_0$ is monotonically
decreasing for galaxies fainter than $L/L^{\ast}<1$ and
$21<\mu_0(B_J)<22$ mag arcsec$^{-2}$ otherwise (see Figure \ref{fig:mods}).
We include passive luminosity evolution in all three models, as the
no--evolution models have been clearly ruled out by both number counts
and our own EBL results.  As Model A (Model PE) has a broader
(narrower) surface brightness distribution than is found by recent LSB
surveys (Sprayberry \etal 1997, Dalcanton \etal 1997, O'Neil \& Bothun
2000), these models are taken as illustrative limits on the fraction
of low surface brightness galaxies in the local universe.  Recent
determinations of the number density of galaxies as a function of both
luminosity and surface brightness (c.f.  Blanton \etal 2001 and Cross
\etal 2001) are well bracketed by these models:  Model A allows for
too many low surface brightness galaxies, while the PE model
clearly allows for too few (see Figure \ref{fig:mods}).  

As described in Table 2 of FM95, each surface brightness distribution
model has been paired with a tuned luminosity function, so that each
model matches the observed morphological distributions and luminosity
functions recovered by local redshift surveys. The models include identical
distributions in the relative number of galaxies of different Hubble
types (E/S0, S0, Sab, Sbc, and Sdm), which are described by
bulge-to-disk flux ratios of 1.0, 0.4, 0.3, 0.15, and 0.0, with small
scatter.  The bulge components for all galaxies have E-type SEDs, and
S0 to Sdm galaxies have disk components with E, Sa, Sb, and Sc-type
SEDs, for which we have used the Poggianti (1997) models.  Bulges were
given r$^{1/4}$--law light profiles with central surface brightnesses
drawn from the empirical relationship found by Sandage \& Perelmuter
(1990), $\mu_0 = -0.48 M_{B_T}+11.02$.  For disk components, we
adopted exponential light profiles, with surface brightnesses drawn
from the 3 model distributions for disk galaxies listed above.

We have calculated passive evolution and K--corrections from the
population synthesis models and SEDs of Poggianti (1997), shown in
Figures \ref{fig:k} and \ref{fig:e}, and we have assumed uniform
comoving density as a function of redshift in all cases.  All models
were run with $H_0=50$ and 70 km s$^{-1}$ Mpc$^{-1}$ and
($\Omega_M=0.2,\Omega_\Lambda=0$), (1,0), and (0.2, 0.8).  Different
values of $H_0$ have little effect ($<10$\%) on the integrated counts
or background.  The total background increases for models with larger
volume --- (1,0), (0.2,0), and (0.2,0.8), in order of increasing
volume --- but the fractional flux as a function of redshift changes
by less than 10\% with cosmological model.  

All three models under-predict the number counts and integrated flux in
observed sources, as expected, and will clearly under-predict the total
EBL as illustrated in the passive evolution model discussed in
\S\ref{lumfuncs}.

\subsection{Results}\label{opt.unres.results}

\begin{figure}[t]
\begin{center}
\includegraphics[width=3in,angle=0]{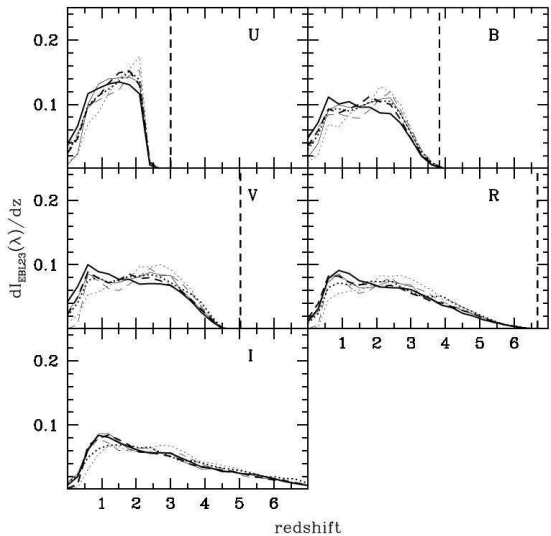}
\caption{\footnotesize
For the Johnson/Cousins bandpasses indicated in each panel and the
models discussed in the text, we plot the redshift distribution of the
EBL --- the differential EBL from all galaxies as a function of
redshift, normalized by the total EBL in each band. Line types
correspond to the models as described in the caption of Figure
\ref{fig:frac.det.with.z}.  In this plot, cosmological models
are virtually indistinguishable because the fractional volume per
redshift bin changes very little with $\Omega$. The dashed vertical
line in each plot indicates the redshift corresponding to the Lyman
limit for the central wavelength of each bandpass.}
\label{fig:ebl.frac.with.z}
\end{center}
\end{figure}

\begin{figure}[t]
\begin{center}
\includegraphics[width=3in,angle=0]{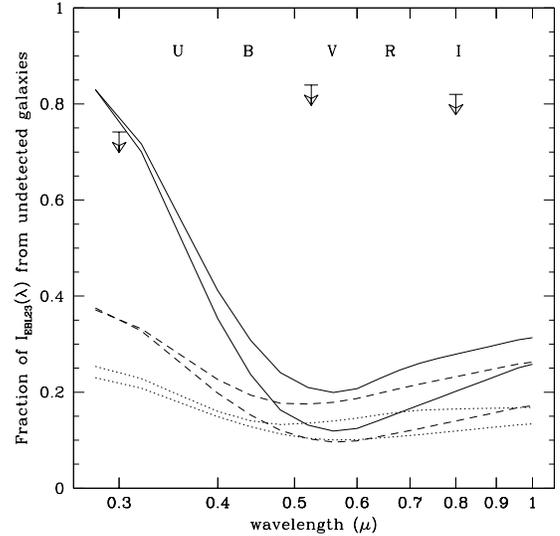}
\caption{\footnotesize The lines show the fraction of the EBL that
comes from undetected galaxies as predicted by our models.  Line types
are as in Figure \ref{fig:frac.det.with.z}.  Arrows show the upper
limits on the fraction of the EBL which might come from undetected
galaxies based on the  EBL detections summarized in \S\ref{sum.detect}
and Table \ref{tab:cum.errors}. These arrows show the ratio of flux
recovered by ensemble photometry (from resolved galaxies) and the two
sigma upper limits of our EBL detection.
See \S\ref{opt.unres.results} for discussion.}
\label{fig:ebl.frac.with.band}
\end{center}
\end{figure}

In Figures \ref{fig:frac.det.with.z} through
\ref{fig:ebl.frac.with.band}, we plot the distribution of the total
flux from the modeled galaxy populations as a function of redshift,
wavelength, and origin (detectable or undetectable galaxies).
Detection limits applied at each bandpass are the $5\sigma$ detection
limits of the HDF catalog (Williams \etal 1996), with appropriate
conversions to the ground--based filter bandpasses, summarized in
Table \ref{tab:model.limits}.  The conversions given in this Table
include differences in the evolutionary corrections and K-corrections
between WFPC2 and UBVRI filters (see Figures
\ref{fig:k}--\ref{fig:eplusk}), which are generally less than 0.3 mag
and change by less than 0.1 mag at $z\gta 0.5$.  We only consider
sources with $V>23$ mag here, and we assume perfect photometry for
sources which meet the detection criteria.

\begin{deluxetable}{c c c | c c c}
\tablewidth{30pc}
\tablecaption{Adopted Detection Limits of the HDF 
\label{tab:model.limits}}
\tablehead{ 
	\colhead{{\sc HST}} & 
	\colhead{$\mu_{\rm core}$} &
	\colhead{$m$} &
	\colhead{Johnson/Cousins} & 
	\colhead{$\mu_{\rm core}$} &
	\colhead{$m$} \\
	\colhead{Filter} &
	\colhead{AB mag arcsec$^-2$} &
	\colhead{AB mag}&
	\colhead{Filter}&
	\colhead{mag arcsec$^-2$}  &
	\colhead{mag}
}
\startdata
{\sc wfpc2/}F300W 	& 25.0	& 27.5	& U & 23.9	& 26.4	\nl 
{\sc wfpc2/}F450W	& 25.8 	& 28.3	& B & 25.8	& 28.3	\nl 
{\sc wfpc2/}F606W 	& 26.3	& 28.8	& V & 25.9 	& 28.4	\nl 
{\sc wfpc2/}F606W	& 26.3	& 28.8	& R & 25.5 	& 28.0	\nl 
{\sc wfpc2/}F814W	& 25.8	& 28.3	& I & 25.2 	& 27.7	\nl 
\enddata
\end{deluxetable}

In Figure \ref{fig:frac.det.with.z}, we show the fraction of the total
flux which comes from undetected sources as a function of redshift.
For all models, this plot demonstrates that if galaxy populations at
higher redshifts are the passively evolving counterparts of those in
the local universe, the flux from undetected sources becomes
significant by redshifts of $1<z<3$.  The undetected fraction is the
highest in the $U$ band, due to the high sky noise and low sensitivity
of the F300W HDF images relative to the other bandpasses which define
our detection criteria. The detection fractions are similar in $B$ and
$V$, where detection limits and galaxy colors are similar.  The
fraction of light from undetected sources in $I$ is small at 
$z<2$ due to the generally red color of galaxies, but
increases beyond that redshift due to cosmological effects.
Model A, with the largest fraction of low $\mu_0$ galaxies,
has the sharpest increase in the undetected EBL with redshift, as
expected.  A balance between evolutionary-- and K--corrections at
$1<z<3$ slow this trend and cause the dip in the fraction of undetected
light in $B$, $V$, and $R$.  The Lyman limit for each band obviously
represents the highest redshift from which one could expect to detect
flux.

In Figure \ref{fig:ebl.frac.with.z}, we plot the distribution of light
with redshift in these models.  All three models have roughly the same
distribution of $I_{\rm EBL}(\lambda,z)$ simply because all models
employ a uniform comoving number density with redshift and the same
passive luminosity evolution.  Although we do not intend to
realistically predict the redshift distribution of the EBL, we show
this plot for comparison with Figure \ref{fig:frac.det.with.z} to
indicate the redshifts at which the majority of undetected galaxies
lie in these models.  Looking at Figures \ref{fig:frac.det.with.z} and
\ref{fig:ebl.frac.with.z} together, it is clear that while $40-100$\%
of the $B$--band flux from $z>3$ is in undetectable sources for all
of the models considered, only a small fraction of the total $B$--band
EBL comes from those redshifts.  Thus, the majority of the light from
unresolved sources comes from $1<z<3$ at $B$ for local-type galaxy
populations in this scenario.

Figure \ref{fig:ebl.frac.with.band} shows the fraction of EBL23 coming
from undetected sources as a function of wavelength.  These models
indicate that 10--35\% of the light from the high redshift
counterparts of local galaxy populations would come from
(individually) undetected sources in bandpasses between $V$ and $I$
with sensitivity limits similar to the HDF, 15--40\% would come from
undetected sources at $B$, and 20--70\% would come from undetected
sources at $U$.  This trend with wavelength (smallest fraction of
undetected sources around 5000\AA) follows the trend in the detection
limits of the HDF bandpasses, as discussed in \S\ref{numcnts}.  Note
that the color of the EBL23 is similar to the color of detected
galaxies (see Figure \ref{fig:eblsummary}) in $V$ and $I$, as is the
$2\sigma$ lower limit of minEBL23 (see also Table
\ref{tab:cum.errors}).

We stress again that cosmological surface brightness dimming and the
fraction of LSBs in each model are the dominant effects which govern
how much light comes from undetected sources and these effects are
independent of wavelength.  The passive luminosity evolution
corrections, K--corrections, and the HDF-specific detection limits we
adopt will determine how the fraction of undetected sources varies
with wavelength.  Finally, we note that although the surface
brightness distribution of galaxies as a function of redshift is
presently unconstrained, and may or may not show significant variation
with redshift, it is unlikely that the surface brightness distribution
at any redshift is significantly more extreme than the distribution
bracketed by our models.  Bearing these uncertainties in mind, we can
use the results of these models to estimate the value of EBL23 based
on the minEBL23 (the flux in individually detected galaxies from
ensemble aperture photometry) and the undetected fractions summarized
above.  If the universe is populated by galaxies with surface
brightness distributions like those in the local universe, then these
models suggest the following values for EBL23: 2.6--7.0\tto{-9} cgs,
1.0--1.3\tto{-9} cgs, and 0.9--1.2\tto{-9} cgs at $U_{300}$, $V_{555}$
and $I_{814}$, respectively.  These ranges are in good agreement with
our detected values for EBL23 (see Table \ref{tab:cum.errors}), and
with the estimates of the EBL23 based on the corrected number counts
we presented in \S\ref{numcnts}.

\section{The Bolometric EBL (0.1--1000\micron)}\label{bolom}

\begin{figure}[t]
\begin{center}
\includegraphics[width=3in,angle=0]{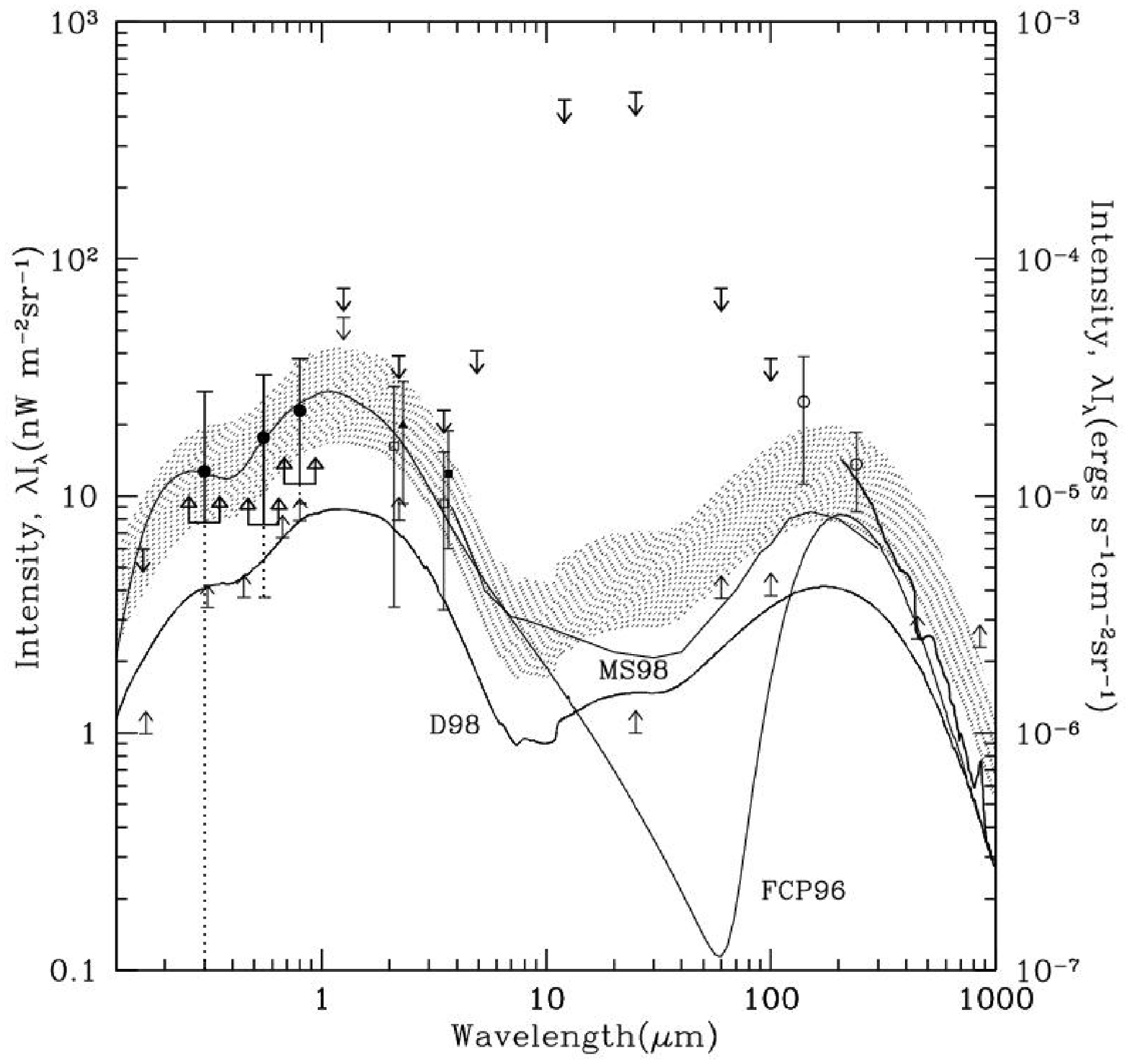}
\caption{\footnotesize EBL detections, limits, and models as a
function of wavelength.  The filed circles show the EBL detections
with $2\sigma$ error bars and lower limit symbols as defined in Figure
\ref{fig:lumden}.  Also plotted are lower limits from Armand \etal 
(1994) at 2000\AA, the HDF (Williams \etal  1996), Gardner \etal
(1997) at 2.2\micron, IRAS (Hacking \& Soifer 1991) at 10--100\micron,
and Blain \etal  (1999) at 450 and 850 \micron.  These lower limits
are based on the integrated flux in detected sources at each
wavelength.  Upper limits marked in bold are from Hurwitz, Bowyer, \&
Martin (1991) at 1600\AA\ and DIRBE (Hauser \etal  1998).  The open
circles indicate DIRBE detections. The bold line at 125--1000\micron\
shows the FIRAS detection (Fixsen \etal  1998). Detections in the
near--IR are from Wright (2000, filled triangle), Gorjian, Wright, \&
Chary (2000, open squares), and Wright \& Reese (2000, filled square).
The lines which indicate models are all labeled and are from Malkan \&
Stecker (1998, MS98), Dwek \etal  (1998, D98), and Fall, Charlot, \&
Pei (1996, FCP96) as described in \S\ref{bolom}. The shaded region
shows the D98 model rescaled to match the range allowed by our EBL
detections.}
\label{fig:models}
\end{center}
\end{figure}

In Figure \ref{fig:models}, we plot EBL detections to date, together
with the integrated light in detectable sources (lower limits to the
EBL) in units of $\nu I_\nu$ between 0.1 and 1000\micron.\footnote{
The total energy per unit increment of wavelength is given by $I =
\int{I_\lambda d\lambda} = \int{\lambda I_\lambda d\ln \lambda}$. 
By plotting energy as $\lambda I_\lambda = \nu I_\nu$ against $\log
\lambda$, the total energy contained in the spectrum as a function of
wavelength is proportional to the area under the curve.  We give
$\lambda I_\lambda$ in standard kms units of  nW m$^{-2}$sr$^{-1}$, 
equivalent to $10^{-6}$ ergs s$^{-1}$cm$^{-2}$sr$^{-1}$.}
The DIRBE and FIRAS detections at $\lambda>100$\micron\ and the
lower limit from IRAS detected galaxies at 10-100\micron\  indicate 
that energy is contained in the far infrared portion of the spectrum.
Given that light from stellar nucleosynthesis is emitted at wavelengths
$0.1-10$\micron, Figure \ref{fig:models} emphasizes the fact that 30\%
or more of the light from stellar nucleosynthesis has been
redistributed into the wavelength range $10-1000$\micron\ by dust
absorption and re-radiation and, to a lesser degree, by cosmological
redshifting. Realistic estimates of the total energy from stellar
nucleosynthesis must therefore be based on the bolometric EBL from the
UV to IR.  In lieu of accurate measurements in the mid--IR range,
realistic models of dust obscuration and the dust re--emission
spectrum (dust temperature) are needed. To discuss the 
optical EBL in the context of star formation,  we must therefore
first estimate the bolometric EBL based on the optical EBL detections
presented here and current measurements in the far-IR. We do so
in the following section.

\subsection{Models}\label{bolom.models}

Efforts to predict the intensity and spectrum of the EBL by Partridge
\& Peebles (1967) and Harwit (1970) began with the intent of
constraining cosmology and galaxy evolution. Tinsley (1977, 1978)
developed the first detailed models of the EBL explicitly
incorporating stellar initial mass functions (IMFs), star formation
efficiencies, and stellar evolution.  Most subsequent models of the
EBL have focused on integrated galaxy luminosity functions with
redshift--dependent parameterization, with particular attention paid
to dwarf and low surface brightness galaxies (see discussions in
Guiderdoni \& Rocca-Volmerange 1990, Yoshii \& Takahara 1988, and
V\"ais\"anen 1996). 

More recent efforts have focused on painting a detailed picture of
star formation history and chemical enrichment based on the evolution
of resolved sources.  The evolution of the UV luminosity density can
be measured directly from galaxy redshift surveys (e.g., Lilly \etal 
1996, Treyer \etal  1998, Cowie \etal  1999, Steidel \etal  1999,
Sullivan \etal  2000), from which the star formation rate with redshift,
$\dot{\rho}_\ast(z)$, can be inferred for an assumed stellar IMF.  The
mean properties of QSO absorption systems with redshift can also be
used to infer $\dot{\rho}_\ast(z)$, either based on the decrease in
\ion{H}{i} column density with decreasing redshift (under the
assumption that the disappearing \ion{H}{i} is being converted into
stars) or based on the evolution in metal abundance for an assumed IMF
and corresponding metal yield (e.g., Pettini \etal  1994; Lanzetta,
Wolfe, \& Turnsheck 1995; Pei \& Fall 1995).  Estimates of the star
formation rate at high redshift have also come from estimates of the
flux required to produce the proximity effect around quasars (e.g.,
Gunn \& Peterson 1965, Tinsley 1972, Miralda-Escude \& Ostriker 1990).
Using these constraints, the full spectrum of the EBL can then be
predicted from the integrated flux of the stellar populations over
time.

Unfortunately, all methods for estimating $\dot{\rho}_\ast(z)$ contain
significant uncertainties.  The star formation rate deduced from the
rest-frame UV luminosity density is very sensitive to the fraction of
high mass stars in the stellar initial mass function (IMF) and can
vary by factors of 2-3 depending on the value chosen for the low mass
cut off (see Leitherer 1999, Meader 1992).  Aside from the large
uncertainties in the measured UV luminosity density due to
incompleteness, resolved sources at high redshift are biased towards
objects with dense star formation and may therefore paint an
incomplete picture of the high-$z$ universe.  Also, large corrections
for extinction due to dust must be applied to convert an observed UV
luminosity density into a star formation rate (Calzetti 1997).

The SFR inferred from QSO absorption systems, whether from consumption
of \ion{H}{i} or increasing metal abundance, is also subject to a
number of uncertainties.  In all cases, samples may be biased against
the systems with the most star formation, dust, and metals: dusty
foreground absorbers will obscure background QSOs, making the
foreground systems more difficult to study.  In addition, large scale
outflows, a common feature of low--redshift starburst galaxies, have
recently been identified in the high--redshift rapidly star--forming
Lyman break galaxies (Pettini \etal  2000), suggesting that changes
in the apparent gas and metal content of such systems with redshift
may not have a simple relationship to $\dot\rho_{\ast}(z)$ and the
metal production rate.  The mass loss rate in one such galaxy appears
to be as large as the star formation rate, and the recent evidence for
\ion{C}{iv} in Ly-$\alpha$ forest systems with very low \ion{H}{i}
column densities ($\lesssim$\tto{14} cm$^{-2}$) suggests that dilution
of metals over large volumes may cause underestimates in the apparent
star formation rate derived from absorption line studies (see Ellison et
al.\ 1999, Pagel 1999, Pettini 1999 and references therein).

Finally, regarding the predicted spectrum of the EBL, the efforts of
Fall, Charlot, \& Pei (1996) and Pei, Fall, \& Hauser (1998) emphasize
the need for a realistic distribution of dust temperatures in order to
obtain a realistic near-IR spectrum.

With these considerations in mind, we have adopted an empirically
motivated model of the spectral shape of the EBL from Dwek
\etal  (1998, D98). This model is based on $\dot{\rho}_\ast(z)$ as deduced
from UV--optical redshift surveys and includes explicit corrections
for dust extinction and re--radiation based on empirical estimates of 
extinction and dust temperature distributions at $z=0$.  The
comoving luminosity density can then be expressed explicitly as the
sum of the unattenuated stellar emission, $\epsilon_s(\nu,z)$,
and the dust emission per unit comoving volume, $\epsilon_d(\nu,z)$.
Equation~\ref{eq:vol.int} then becomes
\begin{equation}\label{eq:int.ebl.epsilon}
I(\lambda,0) = \int_{0}^{z}
{ [\epsilon_s(\nu,z)+\epsilon_d(\nu,z)] dV_c(z) \over 4\pi D_L(z)^2 }
\quad. 
\end{equation}
D98 estimate the ratio $\epsilon_d(\lambda,0)/\epsilon_s(\lambda,0)$
by comparing the UV--optical luminosity functions of optically
detected galaxies with IR luminosity function of IRAS selected
sources. Using values of ${\cal L} =(1.30\pm0.7)\times 10^8 L_\odot$
Mpc$^{-3}$ for the local stellar luminosity density at
0.1--10\micron\ and ${\cal L}=0.53\times 10^8 L_\odot$ Mpc$^{-3}$ for
the integrated luminosity density of IRAS sources, Dwek \etal obtain
$\epsilon_d(\lambda,0)$/$\epsilon_s(\lambda,0)\sim 0.3$.
The redshift independent dust opacity is assumed to be an average
Galactic interstellar extinction law normalized at the $V$--band to
match this observed extinction.  D98 then calculate the EBL spectrum
using the UV-optical observed $\dot{\rho}_\ast(z)$, a
Salpeter IMF ($0.1<M<120 M_\odot$), stellar evolutionary tracks from
Bressan \etal  (1993), Kurucz stellar atmosphere models for solar
metallicity, redshift--independent dust extinction, and dust
re-emission matching the SED of IRAS galaxies.

The starting-point UV-optical $\dot{\rho}_\ast(z)$ for this model is
taken from Madau, Pozzetti, \& Dickinson (1998), which under-predicts
the detected optical EBL presented in Paper I (see \S\ref{opt.disc}).
While D98 discuss two models which include additional star formation
at $z\gtrsim 1$, the additional mass is all in the form of massive
stars which radiate instantaneously and are entirely dust-obscured,
resulting in an {\it ad hoc} boost to the far IR-EBL.  We instead
simply scale the initial Dwek \etal  model by $\times2.2$ to match
the $2\sigma$ lower limit of our EBL detections and $\times4.7$ to
match the $2\sigma$ upper limit, in order to preserve the consistency
of the D98 model with the observed spectral energy density at $z=0$.
In that any emission from $z>1$ will have a redder spectrum than the
mean EBL, simply scaling in this way will produce a spectrum which is
too blue. However, as discussed in \S\ref{lumfuncs}, it is also
possible that the $z<0.5$ UV luminosity density has been
underestimated by optical surveys, so that the bluer spectrum we have
adopted may be appropriate. Note that the resulting model is in
excellent agreement with recent near-IR results at 2.2 and 3.5\micron\
(Wright \& Reese 2000; Gorjian, Wright, \& Chary 2000; Wright 2000)
and also with the DIRBE and FIRAS results in the far-IR. Adopting this
model, we estimate that the total bolometric EBL is $100 \pm 20$
nW~m$^{-2}$sr$^{-1}$, where errors are $1\sigma$ errors associated
with the fit of that template to the data.

Due to the corrections which account for the redistribution by dust of
energy into the IR portion of the EBL, the star formation rate implied
by the unscaled (or scaled) D98 model is 1.5 (or 3.3--7.1) times
larger than the star formation rate adopted by Madau \etal  (1998).
The dust corrections used by Steidel \etal  (1999) produce a star
formation rate which is roughly 3 times larger than used in the
unscaled D98 model, slightly smaller than the scaling range
adopted here, which is consistent with the fact that the CFRS
and Steidel \etal (1999) luminosity densities are slightly
below our minimum values for the EBL, as discussed in
\S\ref{lumfuncs}.

\subsection{Energy from Accretion}\label{agn}

As mentioned briefly in \S\ref{intro}, another significant source of
energy at UV to far--IR wavelengths is accretion onto black holes
in AGN and quasars.  The total bolometric flux from accretion can be
estimated from the local mass function of black holes at
the centers of galaxies for an assumed radiation efficiency
and total accreted mass. Recent surveys find $M_{\rm bh} \approx 0.005
M_{\rm sph}$, in which $M_{\rm sph}$ is the mass of the surrounding
spheroid and $M_{\rm bh}$ is the mass of the central black hole
(Richstone \etal  1998, Magorrian \etal  1998, Salucci \etal  1999, and
van der Marel 1999).  Following Fabian \& Iwasawa (1999), the
energy density in the universe from accretion is given by
\begin{equation}
{\cal E}_{\rm bh} = 0.005 \Omega_{\rm sph}\rho_{\rm crit} c^2 
		    {\eta_{\rm bh} \over (1+z_e)} \quad, 
\end{equation}
in which $\eta_{\rm bh}$ is the radiation efficiency, $ \Omega_{\rm
sph}$ is the observed mass density in spheroids in units of the
critical density, $\rho_{\rm crit}$, and $(1+z_e)$ compensates for the
energy lost due to cosmic expansion since the emission redshift
$z_e$.  The bolometric flux from accretion is then
\begin{equation}
I^{\rm bol}_{\rm bh} = {c\over 4\pi} 
		{{\cal E}_{\rm bh}\over (1+z_e)} \sim 10\, h\,
		{\rm  nW m^{-2} sr^{-1} }
\end{equation}
for $\eta_{\rm bh}\sim 0.1$, $z_e \sim 2$, 
$\rho_{\rm crit}= 2.775\times 10^{11}\, h^2$ M$_\odot$ Mpc$^{-3}$,
$H_0= 100h$ km s$^{-1}$ Mpc$^{-1}$,
and $\Omega_{\rm sph} \sim 0.0018^{+.0012}_{-0.00085}h^{-1}$ 
(Fukugita, Hogan, \& Peebles 1998, FHP98).

The observed X-ray background (0.1-60 keV) is $\sim0.2$~nW m$^{-2}$
s$^{-1}$.  The large discrepancy between the detected X-ray flux and
the estimated flux from accretion has led to suggestions that 85\% of
the energy estimated to be generated from accretion takes place in
dust-obscured AGN and is emitted in the thermal IR (see discussions in
Fabian 1999).  Further support for this view comes from the fact that
most of the soft X-ray background (below 2 keV) is resolved into
unobscured sources (i.e., optically bright quasars), while most of the
hard X-ray background is associated with highly obscured sources
(Mushotzky \etal 2000).  Photoelectric absorption can naturally
account for the selective obscuration of the soft X-ray spectrum.
Best estimates for the fraction of the far-IR EBL which can be
attributed to AGN are then $<10h$ nW m$^{-2}$ sr$^{-1}$, or $<30$\% of
the observed IR EBL.  This is in good agreement with estimates of the
flux from a growing central black hole relative to the flux from stars
in the spheroid based on arguments for termination of both black hole
accretion and star formation through wind-driven ejection of cool gas
in the spheroid (Silk \& Rees 1998, Fabian 1999, Blandford 1999).
Together, these studies suggest that $<15$\% of the bolometric EBL
comes from accretion onto central black holes. 

\section{Stellar Nucleosynthesis: $\Omega_\ast$ and $|Z|$.}\label{stel}

The bolometric flux of the EBL derived in \S\ref{bolom.models} is a
record of the total energy produced in stellar nucleosynthesis in the
universe, and so can be used to constrain estimates of the baryonic
mass which has been processed through stars.  The relationship between
processed mass and background flux depends strongly on the redshift
dependence of star formation and on the stellar IMF, but is only
weakly dependent on the assumed cosmology for the reasons discussed in
\S\ref{lumfuncs} and \S\ref{opt.unres}.

As an illustrative case, we can obtain a simple estimate of
the total mass processed by stars by assuming that all stars formed in
a single burst at an effective redshift $z_e$, and that all the energy
from that burst was emitted instantaneously.  The assumption of
instantaneous emission does not strongly affect the result because
most of the light from a stellar population is emitted by hot,
short--lived stars in the first $\sim10$Myr.  The integrated EBL at
$z=0$ in Equation~\ref{eq:vol.int} then simplifies to
\begin{equation}\label{eq:burst}
I^{\rm bol}_{\ast} = { c \over 4\pi } {{\cal E}_\ast \over (1+z_e)} \quad , 
\end{equation}
in which ${\cal E}_{\ast}$ is the bolometric energy density from
stellar nucleosynthesis and $(1+z_e)$ compensates for energy lost to
cosmic expansion.  In the case of instantaneous formation and
emission, ${\cal E}_{\ast}$ can be expressed in terms of the total
energy released in the nucleosynthesis of He and heavier elements:
\begin{equation} \label{eq:eps}
{\cal E}_{\ast} = {\Omega_\ast \rho_{\rm crit} c^2 \eta 
(\Delta Y\ +\ Z)} \quad,
\end{equation}
in which $\eta$ ($\sim 0.0075$) is the mean conversion efficiency of
energy released in nuclear reactions and $\Delta Y$ and $Z$ are the
mass fractions of $^4$He and metals.  Inverting Equation
\ref{eq:burst}, the total baryonic mass processed through stars 
in this model can be derived from a measurement of the bolometric EBL
using the expression:
\begin{equation} \label{eq:omega_star_1}
\Omega_\ast = { 4\pi  (1+z_e)  \over c^3 \eta \rho_{\rm crit}} 
	 	{ I^{\rm bol}_{\ast} \over \langle\Delta Y +Z\rangle }
\quad .
\end{equation}

We can bracket a reasonable range for $\langle\Delta Y + Z\rangle$ by
assuming the  solar value as a lower limit, and the mass weighted average
of the metal conversion fraction in E/S0 and spirals galaxies as the
upper limit.\footnote{ 
Solar values of $\Delta Y$ and $Z$ are 0.04 and 0.02, implying
$DY/DZ=2$.  Interstellar absorption measurements of $DY/DZ$ in the
solar neighborhood are closer to the range 3--4, implying $\Delta Y
\sim 0.07$.  Helium white dwarfs contribute an
additional 10\% of the local stellar mass to the estimate of $\Delta
Y$ (Fleming, Liebert \& Green 1986), so that we have $\langle\Delta Y
+ Z\rangle = 0.07+0.02+0.1\sim 0.2$ as a local estimate for systems
with solar metallicity. This is similar to estimates for other local
spiral galaxies.  Estimates for E/S0 galaxies are as high as 0.5
(Pagel 1997). (Note that the He mass produced in stars is written as
$\Delta Y$ to distinguish it from the total He mass, which includes a
primordial component.) }
Assuming a 3:2 ratio of E/S0 to Sabc galaxies (Persic \& Salucci
1992), we find $\langle\Delta Y + Z\rangle=0.25\pm0.15$.  For $z_e =
1.5$, the total baryonic mass processed through stars corresponding to
a bolometric EBL of $100 \pm 20$ nW~m$^{-2}$sr$^{-1}$ is then
$\Omega_\ast = 0.0030(\pm0.0019)h^{-2}$ in units of the critical
density, or $0.16(\pm0.10)\Omega_{\rm B}$ for $\Omega_{\rm B}=
0.019(\pm 0.001)h^{-2}$ (Burles \& Tytler 1998).  Again, this
calculation assumes a single redshift for star formation with all
energy radiated instantaneously at the redshift of formation.

The true history of star formation is obviously quite different from
this illustrative case.  For time-dependent emission and formation,
the bolometric EBL is the integral of the comoving luminosity density
corresponding to realistic age-- and redshift--dependent emission (see
Equation \ref{eq:vol.int}). For comparison, instantaneous formation at
the same redshift assumed above ($z_e=1.5$) with a modified Salpeter
IMF and time-dependent emission based on SEDs from Buzzoni (1995)
would imply $\Omega_{\ast}= 0.0037(\pm0.0007)h^{-2}$ for our estimate of
the bolometric EBL (for details see MPD98 and Madau \& Pozzetti 1999,
MP99).  The mean of this estimate is about 20\% higher than that 
from the instantaneous formation and emission model discussed above.  The
two models are very similar because the vast majority of energy from a
stellar population is emitted in the first $\sim10$Myrs.  The quoted
uncertainty is smaller than for our illustrative model because the
error range reflects only the uncertainty in our estimate of the
bolometric EBL and no uncertainties in the adopted IMF.

For the same IMF and SEDs, a redshift--dependent star formation rate
for $0<z<4$ based on the observed UV luminosity density and taking dust
obscuration into account (see Steidel \etal  1999) would imply that
almost twice as much mass is processed through stars than in the
instantaneous formation model above (MP99). 
Relative to the instantaneous--formation models, the same
bolometric EBL flux corresponds to a larger value of
$\Omega_{\ast}$ when we consider time--dependent star formation because
more of the emission occurs at higher redshifts, resulting in greater
energy losses to cosmic expansion.  For our estimate of $I^{\rm
bol}_{\rm EBL}$ and the calculations of MP99 discussed above, we
therefore estimate that total mass fraction processed through stars is
$\Omega_{\ast} = 0.0062(\pm 0.0012)h^{-2}$
or $0.33\pm0.07 \Omega_{\rm B}$.
We adopt this value for the remainder of the paper.

For this value of the total processed mass, we can calculate the
corresponding metal mass which is produced in stellar nucleosynthesis.
To do so requires an estimate of the metal yield --- the mass fraction
of metals returned to the ISM relative to the mass remaining in stars
and stellar remnants.  Best estimates of the metal yield, $y_Z$, lie
between 0.01 (corresponding to a Scalo IMF) and 0.034 (as observed in
the Galactic bulge) (Pagel 1987, 1999).  These values incorporate the full
range predicted by IMF models and observations (see Woosley \& Weaver
1995; Tsujimoto \etal  1995; Pagel \& Tautviaisiene 1997; Pagel
1997).  For $Z_\odot=0.017$ (Grevesse, Noels, \& Sauval 1996), this
metal yield range in solar units is $ y_Z = 1.3\pm0.7 Z_\odot$.  If
the mass fraction remaining in stars and stellar remnants is $f$, then
the predicted metal mass density is given by
\begin{equation}
\Omega_Z = f y_Z \Omega_\ast \quad ,
\end{equation}
which gives 
$\Omega_Z = 0.0040(\pm0.0022) h^{-2} Z_\odot$, or 
$\Omega_Z = 0.24(\pm0.13)Z_\odot \Omega_{\rm B}$, 
for an assumed lock--up fraction of $f=0.5$.

Note that we have assumed that the full flux of the EBL is due to
stellar nucleosynthesis in the above calculations of $\Omega_\ast$ and
$\Omega_Z$. If $\lta 10h$ nW~m$^{-2}$sr$^{-1}$ of the IR EBL is due to
AGN, as estimated in \S\ref{agn}, then the energy emitted by stars is
smaller by about about 7\%, and the inferred mass fractions are then
smaller by about 7\% as well.

\subsection{Comparison with other Observations}\label{compareomega}

The total mass processed by stars is not a directly observable
quantity because some fraction of the processed mass will be hidden in
stellar remnants or recycled back into the ISM.  Estimates of
recycling fraction range from 30--50\% for various IMF models (see
discussions in Pagel 1997), but the cumulative effect of many
generations of star formation and repeated recycling is difficult to
estimate.  Firm lower and upper limits for $\Omega_{\ast}$ are,
however, directly observable: the observed mass in stars and stellar
remnants at $z\sim0$ is a lower limit to the total mass which has been
processed through stars, and the total baryon fraction from Big Bang
nucleosynthesis is an upper limit.  FHP98 estimate the mass fraction
in stars and stellar remnants at $z\sim0$ to be $\Omega_{\rm
stars}=0.0025(\pm0.001) h^{-1}$, corresponding to a mass--to--light
ratio of $(M/L)_B \sim 5.9 (M/L)_\odot$.  In units of $\Omega_{\rm
B}$, this lower limit is $\Omega_{\rm stars}=0.13(\pm 0.05)h
\Omega_{\rm B}$.  Our estimate of the total mass fraction processed
through stars, $\Omega_{\ast} = 0.33(\pm 0.07)\Omega_{\rm B}$, is
comfortably above this lower limit and is, obviously, less than the
upper limit --- the total baryon mass fraction from Big Bang
nucleosynthesis and deuterium measurements.

We can also compare the metal mass fraction predicted by the EBL with
the observed mass fraction in gas, stars, and stellar remnants in the
local universe.  Based on recent estimates by FHP99, $\sim 80$\% of
the observed baryons at $z\sim0$ are located in the intracluster gas
of groups and clusters, 17\% are in stars and stellar remnants, and
only 3\% are in neutral atomic and molecular gas.  The observed metal
mass fraction in hot intracluster gas has been estimated to be at
least $0.33 Z_\odot$ for rich galaxy clusters and $0.25-1
Z_\odot$ in groups based on X-ray observations of Fe features (Renzini
1997, Mushotzky \& Loewenstein 1997). More recent estimates from Buote
(1999, 2000) based on more detailed models of the temperature
distribution of the intracluster gas have found values closer to $1
Z_\odot$ in several clusters and elliptical galaxies.
For a total cluster and group gas mass density of $\Omega_{\rm
gas}=0.011^{+0.013}_{-0.005}h^{-1}$
(FHP98) and assuming a metal
mass fraction of $0.65\pm0.35 Z_\odot$ for clusters and groups of all
masses, the observed metal mass fraction in clusters is
$\Omega_{Z,{\rm gas}} = 0.007^{+0.009}_{-0.005} h^{-1}
 Z_\odot$.
Repeating this exercise for the stellar component, we assume that the
metallicity of stars at $z\sim 0$ is roughly 
solar ($1.0\pm0.25 Z_\odot$)
and that the mass density in stars is $\Omega_{\rm
stars}=0.0025(\pm0.001)h^{-1}$.  The total metal mass in stars and
stellar remnants locally is then $\Omega_{Z,{\rm
star}}=0.0025(\pm0.001) h^{-1}Z_\odot$.  The total metal mass fraction
in the local universe is then $\Omega_Z = \Omega_{Z,{\rm star}} +
\Omega_{Z,{\rm gas}} = 0.0095^{+0.010}_{-0.006}h^{-1} Z_\odot$, or
$0.50^{+0.52}_{-0.32} h Z_\odot \Omega_{\rm B}$.
This estimate is consistent with the value indicated by the bolometric
EBL, $0.24(\pm0.13) Z_\odot \Omega_{\rm B}$, as calculated
above.

\section{Summary and Conclusions}

Based on surface photometry from HST/WFPC2 and simultaneous
ground--based surface spectrophotometry from Las Campanas Observatory,
we find mean values for the flux of the EBL23 (the background
light from sources fainter than $V=23$\ABmag) as follows:
$I_{\rm F300W}= 4.0 \pm 2.5$, 
$I_{\rm F555W}= 2.7 \pm 1.4$, and  
$I_{\rm F814W}= 2.2 \pm 1.0$  in units of \tto{-9} cgs,
where uncertainties quoted are $1\sigma$ combined statistical and
systematic errors.  These results are presented in detail in Paper I
and are summarized in \S\ref{sum.detect}.  Adding in the flux from
sources brighter than $V=23$ \ABmag\ (see Table \ref{tab:cum.errors}),
we find the total EBL flux is
$I_{\rm F300W}= 4.3 \pm 2.6$, 
$I_{\rm F555W}= 3.2 \pm 1.5$, and  
$I_{\rm F814W}= 2.9 \pm 1.1$ cgs.

In the context of these measurements of the EBL, we have discussed
constraints on the slope of number counts, the luminosity density as a
function of redshift, the fraction of galaxies which lie below current
surface brightness detection limits, and the history of
stellar nucleosynthesis and metal production in the universe. 
We reach the following principle conclusions:

\noindent
(1) We find that the corrected number counts at $V$ and $I$ magnitudes
fainter than 23 \ABmag\ obey the relation $N\propto 10^{\alpha m}$
with $\alpha= 0.33\pm0.01$, and $\alpha= 0.34\pm0.01$, respectively,
which is consistent with the slope found at brighter magnitudes
(e.g. Smail \etal 1995, Tyson 1988).  This is significantly steeper
than the slope of the raw HDF number counts ($\sim 0.24\pm 0.1$ at
$V>23$ \ABmag, and $\sim 0.22\pm 0.1$ at $I>23$ \ABmag). In contrast
with the raw counts, the corrected counts show no decrease in slope to
the detection limit.  If we integrate the corrected number counts down
to an apparent magnitude corresponding roughly to a dwarf galaxy
($M_V\sim -10$ mag) at $z\sim3$, $V\sim38$ \ABmag, we obtain a total
flux of $1.2$\tto{-9} cgs in both $V$ and $I$.  This is
$1.2\sigma$ below the mean EBL23 flux we estimate at
$V_{606}$ ($I_{814}$), suggesting that number counts would need to be
steeper over some range in apparent magnitude fainter than the current
detection limits in order to obtain the mean EBL flux we detect,
or that the value of EBL23 is roughly $\sim1\sigma$ below
our mean detections at $V$ and $I$.

\noindent
(2) Based on a local luminosity density consistent with Loveday \etal (1992),
passive evolution in the luminosity density of galaxies under-predicts
the EBL by factors of roughly 3, 2, and 2 at $U_{300}$, $V_{555}$, and
$I_{814}$, respectively.  Note, however, that if the local luminosity
density is a factor of two higher than the Loveday \etal values we
have adopted here, as found by Blanton \etal (2001), then passive
evolution agrees with the flux in resolved galaxies (minEBL23) and
with our mean EBL detections to within $1\sigma$.  The mean detected EBL
therefore requires stronger evolution in the luminosity density than
passive evolution will produce, however, the exact form of that
evolution is not well constrained by our results.  

Adopting the local luminosity density assumed by Lilly \etal (1996,
CFRS), the $1\sigma$ upper limits of the cumulative flux measured by
Lilly \etal from redshifts $0<z<1$ is smaller than the flux in
resolved sources by more than a factor of 2: this fact alone
demonstrates that significant flux must be contributed by galaxies at
redshifts $z>1$.  If we adopt ${\cal L}(\lambda,z)\propto
(1+z)^{\delta(\lambda)}$ for the luminosity density at $0<z<1$ based
on the Lilly \etal results, then constant luminosity density at $z>1$,
such as suggested by Steidel \etal (1999) is consistent with the
detected flux in sources at $V_{555}$ and $I_{814}$, and with the
detected EBL at $U_{300}$.  At the upper limit of the EBL detections,
we find that the luminosity density can continue to rise as a power
law to $z\sim 2.5$ without over--predicting the EBL.
 
\noindent
(3) We have modeled the effects of cosmological K-corrections, passive
evolution, and $(1+z)^4$ cosmological surface brightness dimming on
the detectability of local--type galaxy populations as a function of
redshift. For these models, we have adopted the spatial resolution and
surface brightness limits of the HDF. For models which bracket the
observed surface brightness distribution of galaxies in the local
universe, we find that roughly 10--40\% of the EBL from galaxies
fainter than $V\sim23$ (i.e. those sampled in an HDF--sized image),
comes from galaxies which are, at present, individually undetectable at
wavelengths $\lambda>4500$\AA, and roughly 20--70\% comes from
individually undetected galaxies at $\lambda<4500$\AA. Most of the
flux from a local-type galaxy population located at $z=3$ would come
from sources that would not be individually detected in the HDF.
Our models  indicate that the true EBL is likely to be
between the mean detected EBL23 values and the $1\sigma$ 
lower limits of those detections at $V$ and
$I$, and within $\pm 1\sigma$ at $U$.

\noindent
(4) Scaling the model of the bolometric EBL derived by Dwek \etal 
(1998), which is based on a combined UV-optical estimate of the star
formation rate and a model for dust obscuration and re--emission based
on the spectrum of IRAS sources, we find that the optical EBL we
detect corresponds to a total bolometric EBL (0.1 to 1000\micron)
of 100$\pm20$ nW~m$^{-2}$sr$^{-1}$.

\noindent
(5) From this estimate of the total bolometric EBL, we estimate that
the total baryonic mass processed through stars is $\Omega_\ast =
0.0062(\pm0.0012) h^{-2}= 0.33(\pm0.07) \Omega_{\rm B}$, and that the
mean metal mass density in the universe is $\Omega_Z=
0.0040(\pm0.0022) h^{-2} Z_\odot = 0.24(\pm0.13) Z_\odot \Omega_{\rm
B}$, for $\Omega_{\rm B}=0.019(\pm 0.001)h^{-2}$ (Burles \& Tytler
1998).  These estimates are consistent with limits from other
observational constraints.

\acknowledgments

We would like to thank the referee, M. Bershady, for detailed and
helpful comments.  We also thank R.\ Carlberg, J.\ Dalcanton, E.\
Dwek, P.\ Madau, R.\ Marzke, J.\ X.\ Prochaska, T.\ Small, and I.\
Smail for helpful discussions.  This work was supported by NASA
through grants NAG LTSA 5-3254 and GO-05968.01-94A to WLF, and by NASA
through Hubble Fellowship grant \# HF-01088.01-97A awarded by STScI to
RAB.

\end{document}